\documentclass[twocolumn,english,aps,prb,showpacs,byrevtex,amsmath,amssymb,bm,superscriptaddress]{revtex4-1}
\usepackage[T1]{fontenc}
\usepackage[latin9]{inputenc}
\usepackage{textcomp}
\usepackage{amsmath}
\usepackage{graphicx}

\makeatletter
\usepackage{lipsum}
\usepackage{epsfig}
\usepackage{color}
\usepackage{physics}

\definecolor{Red}{rgb}{1,0,0}
\definecolor{Blu}{rgb}{0,0,1}
\definecolor{Green}{rgb}{0,1,0}

\makeatother

\usepackage{babel}
\begin{document}

\title{Multiple band-crossings and Fermi surface topology:\\
	role of double nonsymmorphic symmetries in MnP-type crystal structures}

\author{Giuseppe Cuono}

\affiliation{Dipartimento di Fisica "E.R. Caianiello", Universit\`a degli Studi di Salerno, I-84084 Fisciano
(SA), Italy}

\affiliation{International Research Centre Magtop, Institute of Physics, Polish Academy of Sciences,
Aleja Lotnik\'ow 32/46, PL-02668 Warsaw, Poland}

\author{Filomena Forte}

\affiliation{Consiglio Nazionale delle Ricerche CNR-SPIN, UOS Salerno, I-84084 Fisciano (Salerno),
Italy}

\affiliation{Dipartimento di Fisica "E.R. Caianiello", Universit\`a degli Studi di Salerno, I-84084 Fisciano
(SA), Italy}

\author{Mario Cuoco}

\affiliation{Consiglio Nazionale delle Ricerche CNR-SPIN, UOS Salerno, I-84084 Fisciano (Salerno),
Italy}

\affiliation{Dipartimento di Fisica "E.R. Caianiello", Universit\`a degli Studi di Salerno, I-84084 Fisciano
(SA), Italy}

\author{Rajibul Islam}
\affiliation{International Research Centre Magtop, Institute of Physics, Polish Academy of Sciences,
Aleja Lotnik\'ow 32/46, PL-02668 Warsaw, Poland}

\author{Jianlin Luo}

\affiliation{Beijing National Laboratory for Condensed Matter Physics and Institute of Physics, Chinese Academy of Sciences,
Beijing 100190, China}

\affiliation{Songshan Lake Materials Laboratory, Dongguan, Guangdong 523808, China}

\affiliation{School of Physical Sciences, University of Chinese Academy of Sciences, Beijing 100190, China}

\author{Canio Noce}
\affiliation{Dipartimento di Fisica "E.R. Caianiello", Universit\`a degli Studi di Salerno, I-84084 Fisciano
(SA), Italy}
\affiliation{Consiglio Nazionale delle Ricerche CNR-SPIN, UOS Salerno, I-84084 Fisciano (Salerno),
Italy}

\author{Carmine Autieri}
\email{autieri@ifpan.edu.pl}
\affiliation{International Research Centre Magtop, Institute of Physics, Polish Academy of Sciences,
Aleja Lotnik\'ow 32/46, PL-02668 Warsaw, Poland}
\affiliation{Consiglio Nazionale delle Ricerche CNR-SPIN, UOS Salerno, I-84084 Fisciano (Salerno),
Italy}

\date{\today}
\begin{abstract}
		
We use relativistic {\it ab-initio} methods combined with model Hamiltonian approaches to analyze the normal-phase electronic and structural properties of the recently discovered WP superconductor. Remarkably, the outcomes of such study can be employed to set fundamental connections among WP and the CrAs and MnP superconductors belonging to the same space group. 
One of the key features of the resulting electronic structure is represented by the occurrence of multiple band crossings along specific high symmetry lines of the Brilloiun zone. In particular, we demonstrate that the eight-fold band degeneracy obtained along the SR path at ($k_{x}$,$k_{y}$)=($\pi$,$\pi$) is due to inversion-time reversal invariance and a pair of nonsymmorphic symmetries. The presence of multiple degenerate Fermi points along the SR direction constraints the topology of the Fermi surface, which manifests distinctive marks when considering its evolution upon band filling variation. 
If the Fermi level crosses the bands along the SR line as it happens at the nominal filling of the MnP, these Fermi surfaces are open or closed Fermi pockets.
Moving the relative position of the Fermi level away from the eight-fold degenerate bands as for the WP and CrAs compounds, the electronic changeover exhibits a simultaneous modification of the Fermi surface dimensionality and topology. Four two-dimensional (2D) Fermi surface sheets are centered around the SR line with a corrugated profile along the $k_{z}$ direction. 
Moreover, we show that the spin-orbit interaction determines a selective removal of the band degeneracy and, consequently, a splitting of the quasi 2D Fermi sheets, as it happens in WP.
Finally, we comment on the connections between our results and recent experimental and theoretical proposals about the triplet superconductivity in this class of compounds. 

\end{abstract}


\maketitle

\section{Introduction}

Recent years have testified an increasing interest in the study of nodal metals with protected band degeneracies near the Fermi level, especially due to the rapid development of the field of topological condensed matter.~\cite{Konig07,Hsieh08,Hasan10,Qi11,Hasan11,
Schnyder15,Senthil15,Chiu13,Chiu14,Chiu16,Ando14,
Shiozaki14,Morimoto13,Chan16,Zhao16b,Brzezicki,Groenendijk,Kruthoff17,Slager13,Bouhon18,Murakami18}
Topological gapless phases are marked by topologically protected Fermi surface which occur as a consequence of band crossings associated with a topological number that in turn is tied to the symmetries enforcing the band degeneracy. 
The significant advancements in the classification schemes based on topology and in materials predictions led to the discovery of a wide number of topological semimetals (SMs) that include both Dirac\cite{Wehling14} and Weyl SMs \cite{Armitage18,Young12,Weng15}, nodal line \cite{Horava05, Heikkila11, Burkov11}, type-I and type-II SMs\cite{Soluyanov15}, multifold SMs \cite{Bradlyn16}. 
\\
It is well known that in crystals degeneracies can also come from non point-group types of symmetries. For instance, nonsymmorphic crystalline symmetries, namely symmetries that involve not only point group operations but also non-primitive lattice translations, typically force bands to cross and can enforce the occurrence of unconventional topological phases.~\cite{Wieder16,Bzdusek16,WangN16,Lu16,Parameswaran13,
Michel01,Konig97,Watanabe16,Zhao16,Singh18,Weng16,Mele19,Armitage18,Wieder16B} 
Nonsymmorphic symmetries when combined with inversion, time or particle-hole symmetry transformations,\cite{Brzezicki17} can allow band degeneracies at the origin or at the boundary of the Brillouin zone (BZ).~\cite{Zhao16} Moreover, such nonsymmorphic nodal systems can also exhibit topologically protected Fermi surfaces of reduced dimensionality, as well as topological response phenomena with non-standard magnetotransport properties.~\cite{Zhao16,Yoshida19,Daido19}


\begin{figure}[h]
	\centering
	\includegraphics[width=\columnwidth, angle=0]{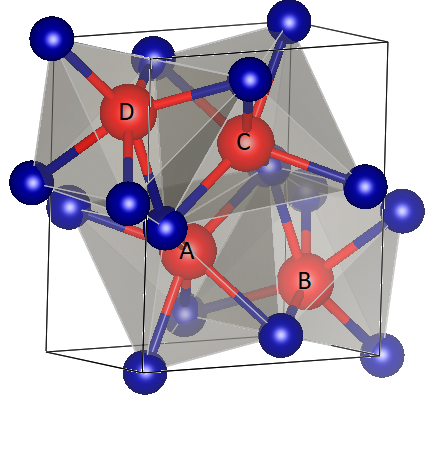}
	\caption{Crystal structure of the WP in the Pnma symmetry after atomic relaxation. W and P atoms are shown as red and blue balls, respectively. The face-shared WP$_6$ octahedra are shown in light gray. The W atoms are labelled as A,B,C,D.}
	\label{Structure}
\end{figure}

\noindent Among the materials exhibiting nonsymmorphic crystal symmetries, a remarkable case is represented by the CrAs, especially in view of the promising possibility to combine non-standard nodal electronic states with magnetic and superconducting orderings.~\cite{Feng15,Liang15,Narayanan15,Wu14}  CrAs belongs to transition metal pnictides with formula MX (M=transition metal, X=P, As, Sb), having orthorhombic MnP-type crystal structure at ambient conditions.~\cite{Chen19,Wang16,Wu10,Wu14,Kotegawa14,Autieri17,Autieri17b,Autieri18}
A member of the same family is the MnP that shares with the CrAs a superconducting phase driven by external pressure and akin to the magnetism, suggesting a new form of superconductivity with non-conventional order parameter.~\cite{Norman11,Goll06,Wu10,Wu14,Kotegawa14,Cheng15,Nigro18} In both compounds the critical temperature-external pressure phase diagram has a typical dome-shaped behaviour ~\cite{Kotegawa14,Kotegawa15,Khasanov15,Keller15,Varma99,Vandermarel03}
with the superconductivity appearing at the critical pressure P equal to 0.7 GPa for CrAs and 8 GPa for MnP, where the helimagnetic transition is suppressed.~\cite{Wu14,Cheng15}\\
Another common and salient feature of the phase diagram of these compounds is the presence of tunable magnetic phases that can coexist with superconductivity,~\cite{Khasanov15} exhibit different types of ordered patterns (e.g. helimagnetism in CrAs~\cite{Wu10,Wu14} and metamagnetism,\cite{Gercsi10,Autieri12}  ferromagnetism, helimagnetism in MnP \cite{Matsuda16}), non-standard magnetic anisotropy~\cite{Keller15,Shen16,Takase79,Yanase80}, or undergo a variety of phase transitions upon cooling\cite{Matsuda16}, doping~\cite{Matsuda18} and application of external pressure.~\cite{Kotegawa15} 

Very recently, Liu et al.~\cite{Liu19} found a new superconductor, namely the WP, belonging to this class of transition metal pnictides, with a bulk superconductivity appearing at 0.84 K, at ambient pressure.~\cite{Liu19}
Remarkably, the WP is the first 5$d$ transition metal phosphide with a non magnetic ground state.
The extended 5$d$-shells lead to a strong coupling between the W $d$-orbitals and the neighboring $p$-orbitals, giving rise to a crystal structure more distorted than that of CrAs and MnP. Moreover, the W-5$d$ electrons exhibit spin-orbit coupling interaction stronger than the 3$d$ of the CrAs and MnP.

In this paper, we employ density functional theory (DFT), supported by the formulation of an effective low-energy model Hamiltonian, to investigate the electronic band-structure of WP, comparing it with the non-magnetic phase of the CrAs and MnP. We demonstrate that the presence of nonsymmorphic crystal symmetries, other than time-reversal and inversion symmetries, allows for additional degeneracies in the band spectrum, along some high-symmetry paths of the BZ. In particular, we show that an extra four-fold degeneracy, beyond time-inversion degeneracy, is due to the invariance of the Hamiltonian with respect to a pair of nonsymmorphic transformations acting along the SR line of the BZ, and we derive the explicit common basis along the high symmetry lines. Then, we analyze the symmetry constraints on the topology of the Fermi surface by considering different band fillings. 
We show that close to the R point, if there are no Fermi points along the SR line, the electronic changeover exhibits a simultaneous modification of the Fermi surface dimensionality and topology forming four two-dimensional (2D) Fermi surface sheets centered around the SR line at ($k_{x}$,$k_{y}$)=($\pi$,$\pi$) with a corrugated profile along the $k_{z}$ direction. Otherwise, the bands along the SR line cross the Fermi level thus contributing to the formation of eight Fermi surfaces that in turn can get converted into Fermi pockets and stripe-shaped Fermi lines.

The paper is organized as follows: in the next section we describe the computational details of the {\it ab-initio} calculations. In section III we compare the structural properties of WP with those of CrAs and MnP, whereas in section IV we present our DFT results for the electronic band structure and related density of states. Section V is devoted to the explicit derivation of the nonsymmorphic symmetry operators and to demonstrate the eight-fold degeneracy along the SR line, while in section VI we calculate the Fermi surfaces for the different compounds. Finally, the last section is devoted to the final remarks and conclusions.

\section{Computational details}

We have performed DFT calculations by using the VASP package.~\cite{Kresse96}
The core and the valence electrons were treated within the Projector Augmented Wave (PAW) method~\cite{Kresse99} with a cutoff of 400 eV for the plane wave basis. We have used a PAW with 6 valence electrons for the W (6s$^2$5d$^4$), 7 valence electrons for the Mn (4s$^1$3d$^6$) and 5 valence electrons for the P (3s$^2$3p$^3$).
The calculations have been performed using a 12$\times$16$\times$10 $k$-point Monkhorst-Pack grid~\cite{Monkhorst76} for the non polarized case and a 6$\times$8$\times$6 $k$-point Monkhorst-Pack grid for the case with spin-orbit coupling (SOC). In the first case we have a 240 $k$-points in the indipendent BZ, while we have 288 $k$-points in the case with SOC.  For the treatment of exchange-correlation, the generalized gradient approximation (GGA)\cite{Perdew96} has been used.
For the CrAs, we have used the computational setup of reference \onlinecite{Autieri17}.

We have optimized the internal degrees of freedom by minimizing the total energy to be less than 7$\times10^{-7}$ eV. After obtaining the Bloch wave functions $\psi_{n,\textbf{k}}$, the Wannier functions~\cite{Marzari97,Souza01} have been built up using the WANNIER90 code~\cite{Mostofi08} generalizing the following formula to get the Wannier functions $W_n(\textbf{r})$:

\begin{equation}
  W_n(\textbf{r})=\frac{V}{(2\pi)^3}\int d\textbf{k} \psi_{n,\textbf{k}}e^{-i\textbf{k}\cdot\textbf{r}}\, ,
\end{equation}
were $V$ is the volume of the unit cell and $n$ is the band index.

To extract the low energy properties of the electronic bands, we have used the Slater-Koster interpolation scheme as implemented in Wannier90. In particular, we have fitted the electronic bands, in order to get the hopping parameters and the spin-orbit constants. This approach has been applied to determine the real space Hamiltonian matrix elements in the maximally localized Wannier function basis, and to find out the Fermi surface with a 50$\times$50$\times$50 $k$-point grid.

\section{Structural properties}

WP belongs to the family of transition-metal pnictides with a general formula MX. Among the various phases exhibited by these compounds, we focus on the orthorhombic MnP-type B31 phase (space group Pnma),~\cite{Wilson64} being the one where the superconducting phase occurs for WP, CrAs and MnP. The primitive cell contains four M and four X atoms. Each M atom is surrounded by six nearest-neighbour X atoms and it is located at the centre of MX$_6$ octahedra, which are face-sharing as shown in Fig.~\ref{Structure}. Four of the six bonds are inequivalent due to the large anisotropy exhibited by this class of compounds. 
In Table I, we compare the structural properties of WP, CrAs and MnP at ambient pressure with the ideal high-symmetry crystal structure of the WP. The lattice constants of the high-symmetry structure are given by: a=$\sqrt[3]{V\sqrt{2}}$, b=$\sqrt[3]{V/2}$, c=$\sqrt[3]{V\sqrt{2}}$, where $V$ is the volume of primitive cell of the WP.
The experimental investigation on WP provided just the lattice constants, without any info about the refinement of the unit cell and the corresponding internal degrees of freedom~\cite{Liu19}. Hence, while the internal degrees of freedom used for the DFT calculation of CrAs and MnP have been assumed from the experimental observations, for the WP we had to explicitly calculate them by means of the atomic relaxations within the DFT.
\\
We notice that $a$ and $c$ of the WP are larger than those of CrAs and MnP; the $b$ lattice constant of CrAs is the largest and it is responsible for the antiferromagnetic transition exhibited by this compound\cite{Wu14}.
In the Table \ref{tab:table1}, we also report the values of the four inequivalent bonds M-X$_{1}$-M-X$_{4}$, and observe that M-X$_{2}$ and M-X$_{3}$ are almost degenerate. Finally, we have computed the polyhedral volume (PV), namely the volume of the octahedra that can be build around the M atoms, and the bond angle variance (BAV) of these octahedra,
which we define as $\sum_{i=1,6}[\frac{(\theta_{i}-\theta_{0})^2}{n-1}]$ where $\theta_{0}$ is the ideal bond angle equal to 90$^{\circ}$ for a regular octahedron.
Concerning the PV, we notice that the value for WP is larger than MnP since the atomic radius of W is larger with respect to Mn. Moreover, the PV of CrAs has the largest value because of the large volume of the As. The estimated BAV values show that the WP exhibits a more distorted structure if compared to CrAs and MnP.

\begin{table}
  \begin{center}
    \caption{Unit-lattice cell parameters, coordinates of the M and X atoms, bond lengths connecting M and X atoms, polyhedral volume (PV) and bond angle variance (BAV), of the transition-metal pnictides MX compared with the ideal high-symmetry structure (HSS). Space group:
Pnma. Atomic positions: M: 4c (x, 1/4, z); X: 4c (x, 1/4, z).}
    \label{tab:table1}
    \begin{tabular}{c|c|c|c|c}
       & \textbf{WP}~\cite{Liu19} & \textbf{CrAs}~\cite{Shen16} & \textbf{MnP}~\cite{Cheng15,Rundqvist62} & \textbf{HSS} \\
      \hline
      a(\AA) & 5.7222 & 5.60499 & 5.236 & 5.4627 \\ 
      b(\AA) & 3.2434 & 3.58827 & 3.181 & 3.8627 \\
      c(\AA) & 6.2110 & 6.13519 & 5.896 & 5.4627 \\ 
     V(\AA $^{3}$) & 115.27 & 123.392 & 98.20 & 115.27 \\
      \hline
     $ x_{M}$ & 0.0138{*}  & 0.0070 & 0.0049 & 0 \\
      $z_{M}$ & 0.1880{*} & 0.2049 & 0.1965 & 1/4\\
     $ x_{X}$ & 0.1822{*} & 0.2045 & 0.1878 & 1/4 \\
     $ z_{X}$ & 0.5654{*} & 0.5836 & 0.5686 & 1/2\\
     \hline
     Bond length(\AA)& & & &\\
     M-X$_{1}$ & 2.5342{*} & 2.5736 & 2.3938 & 1.9314 \\
     M-X$_{2}$ & 2.4967{*} & 2.5116 & 2.3379 & 2.7314 \\
     M-X$_{3}$ & 2.4972{*} & 2.5274 & 2.3848 & 2.7314\\
     M-X$_{4}$ & 2.4653{*} & 2.4511 & 2.2803 & 1.9314 \\
     \hline
     PV(\AA$^{3}$) & 19.212{*} & 20.565 & 16.367 & 19.212  \\ 
     BAV (deg.$^{2}$) & 191.49{*} & 77.708 & 142.98 & 0 \\
     \hline
    \end{tabular}
  \end{center}
 \begin{flushleft} 
  * present work
   \end{flushleft} 
\end{table}

\section{Electronic properties}
In this section, we present the DFT based results of the electronic properties of WP. We start by considering the DOS computed for the experimental crystal structure discussed in the previous section, in order to extract the dominant orbital character in the different regions of the energy spectrum. Then, we present the electronic band structure using the crystal parameters reported in Table \ref{tab:table1}.

\subsection{Density of states}

Here, we investigate the density of states (DOS) obtained by using the experimental lattice constants and the atomic positions calculated from the atomic relaxation procedure as in Table \ref{tab:table1}.

\begin{figure}[t!]
\centering
\includegraphics[height=\columnwidth, angle=270]{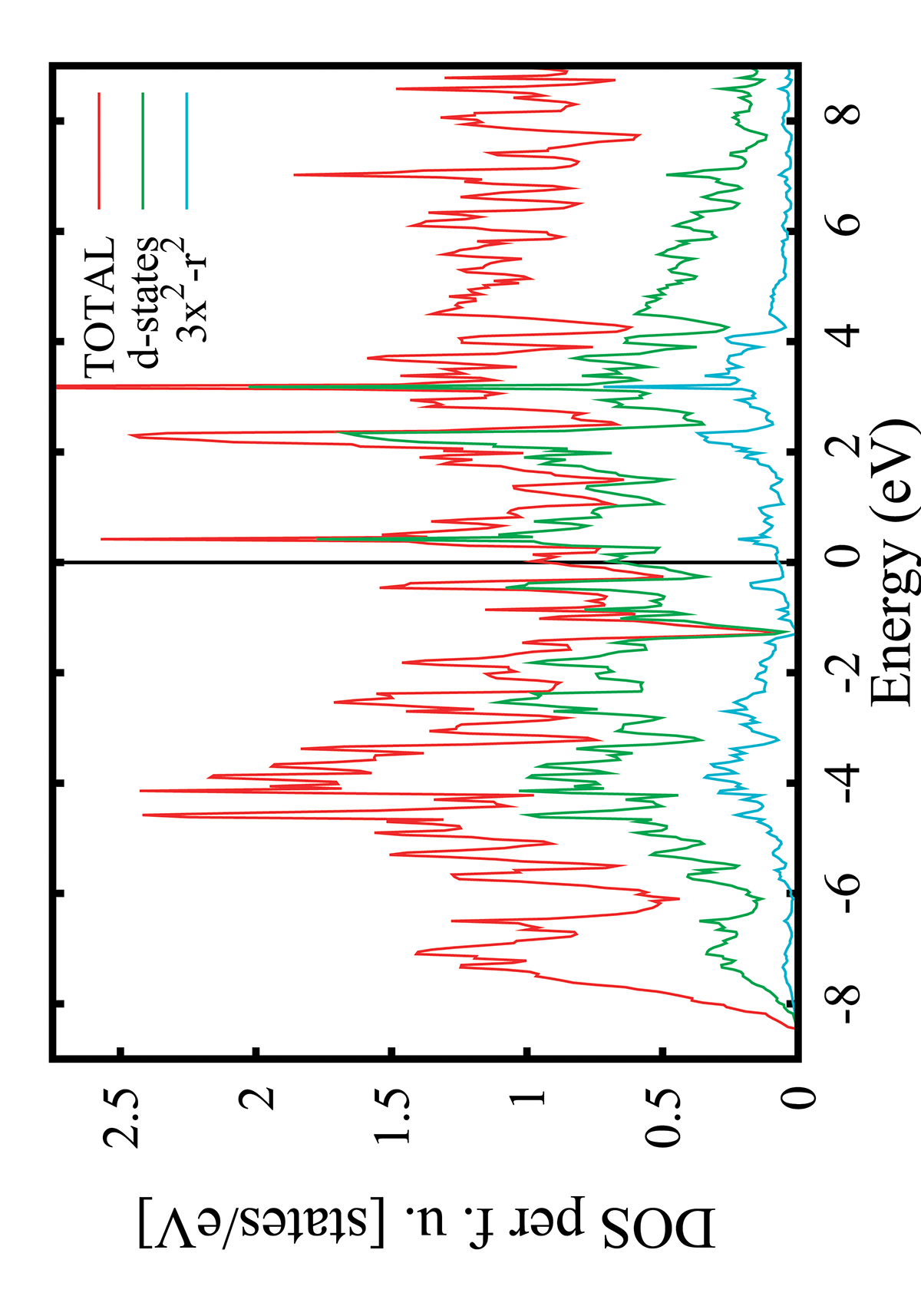}
\caption{DOS of the WP by using the experimental lattice constants and the internal positions calculated in this work. The total DOS per formula unit is plotted as red.
W states are plotted as green and the $3x^{2}-r^{2}$ states are shown in cyan.
The Fermi level is set at the zero energy.
}
\label{DOS3}
\end{figure}

\noindent As we can see from Fig.~\ref{DOS3}, referring to the character of the bands, we infer that the bands close to the Fermi level are primarily due to the W degrees of freedom, so they are more flat with respect to the bands that are located 4~eV above and below the Fermi level, and this leads to peaks in the DOS.
The latter bands may assigned to P states, and are more delocalized with respect to the W ones. 

\begin{figure}[]
\centering
\includegraphics[height=\columnwidth, angle=270]{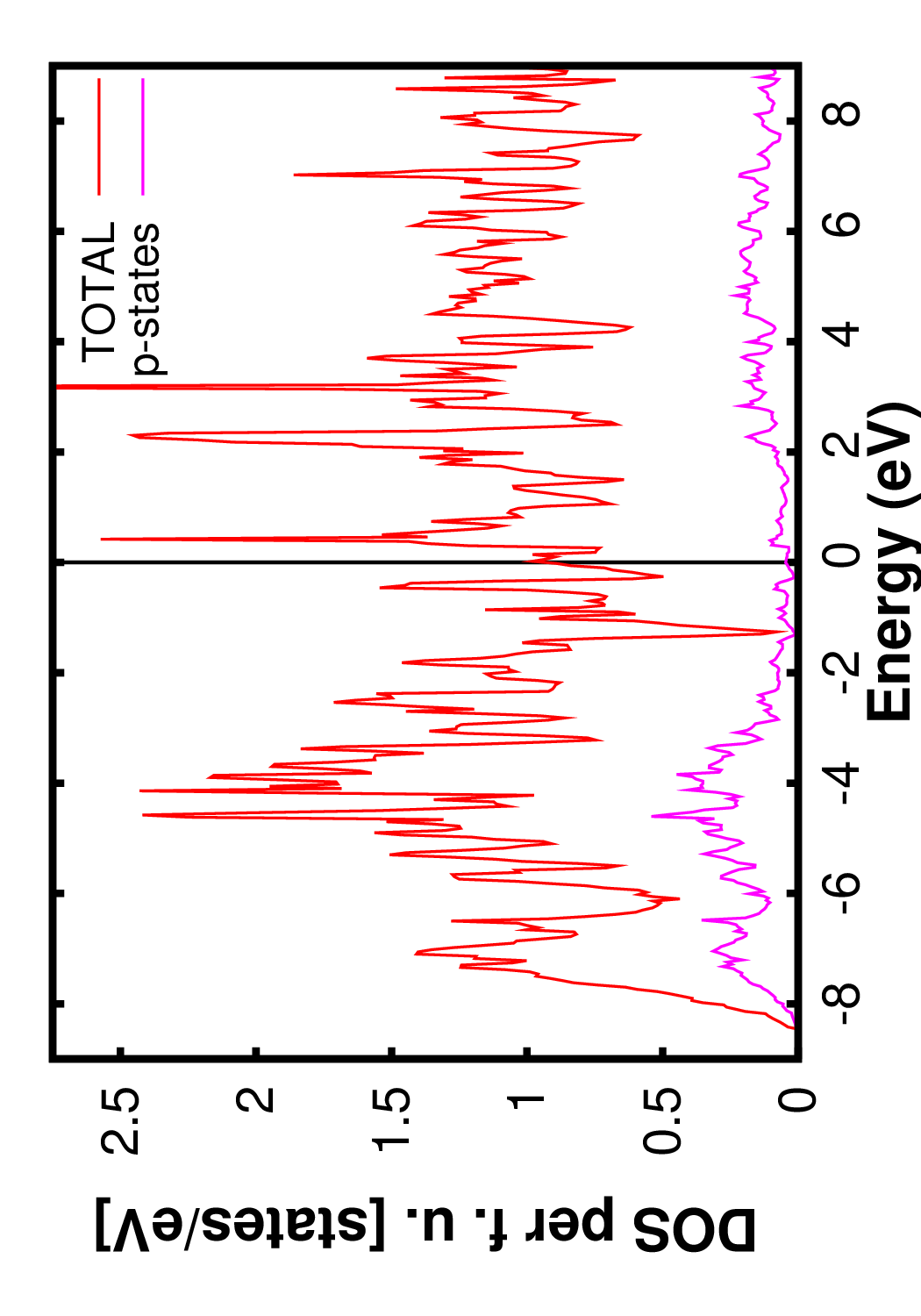}
\caption{DOS of the WP by using the experimental lattice constants and the internal positions calculated in this work. The total DOS per formula unit is plotted as red and the P-3$p$ states are plotted as magenta.
The Fermi level is set at the zero energy.
}
\label{DOS4}
\end{figure}

The character of the DOS from -8.5~eV to -6~eV is predominantly P-3$p$, while from -6~eV to 4~eV the W-5$d$ states dominate, and finally, above 4~eV there is a mixing between P-3$p$, P-4$s$ and W-6$s$ states.
For completeness, in Fig.~\ref{DOS4} we show the P-3$p$ states together with the total DOS by using the experimental lattice constants and the internal positions calculated within the DFT.

We remark that, since the W bands are due to 5$d$ orbitals, the bandwidth is much wider than that of CrAs and MnP, so that we may suppose that the WP is the least correlated system between the three transition metal pnictides investigated.
In the Appendix A we analyze the effect of the octahedral rotations on the crystal-field energy levels by looking at the density of states of WP in three cases.
As for CrAs,~\cite{Autieri17} it is hopeless to entirely decouple the W-5$d$ states close to the Fermi level from the P-3$p$ states in an accurate way, because of the strong hybridization between them, as we show in the Appendix B.
Nevertheless, an accurate effective model for a reduced number of $d$-bands can be obtained also in the present case of strong hybridization.~\cite{Cuono19,Cuono18AIP} 
In the Appendix C we comment about the oxidation state of CrAs, WP and MnP.

\subsection{\textcolor{black}{Band structure and effect of the spin-orbit coupling on the band degeneracy}}

\begin{figure}
	\centering
	\includegraphics[width=6.3cm,angle=270]{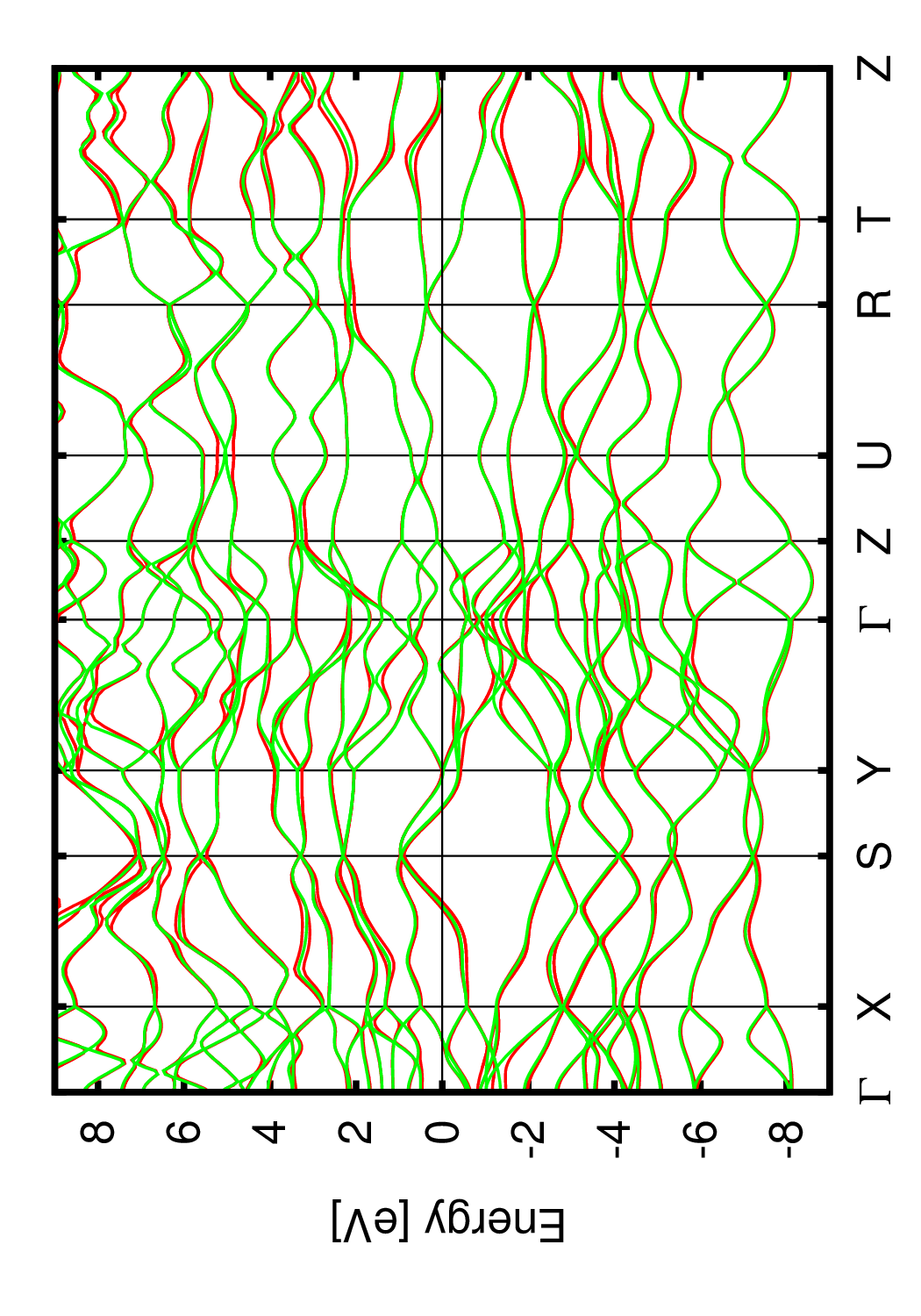}
	\includegraphics[width=6.3cm,angle=270]{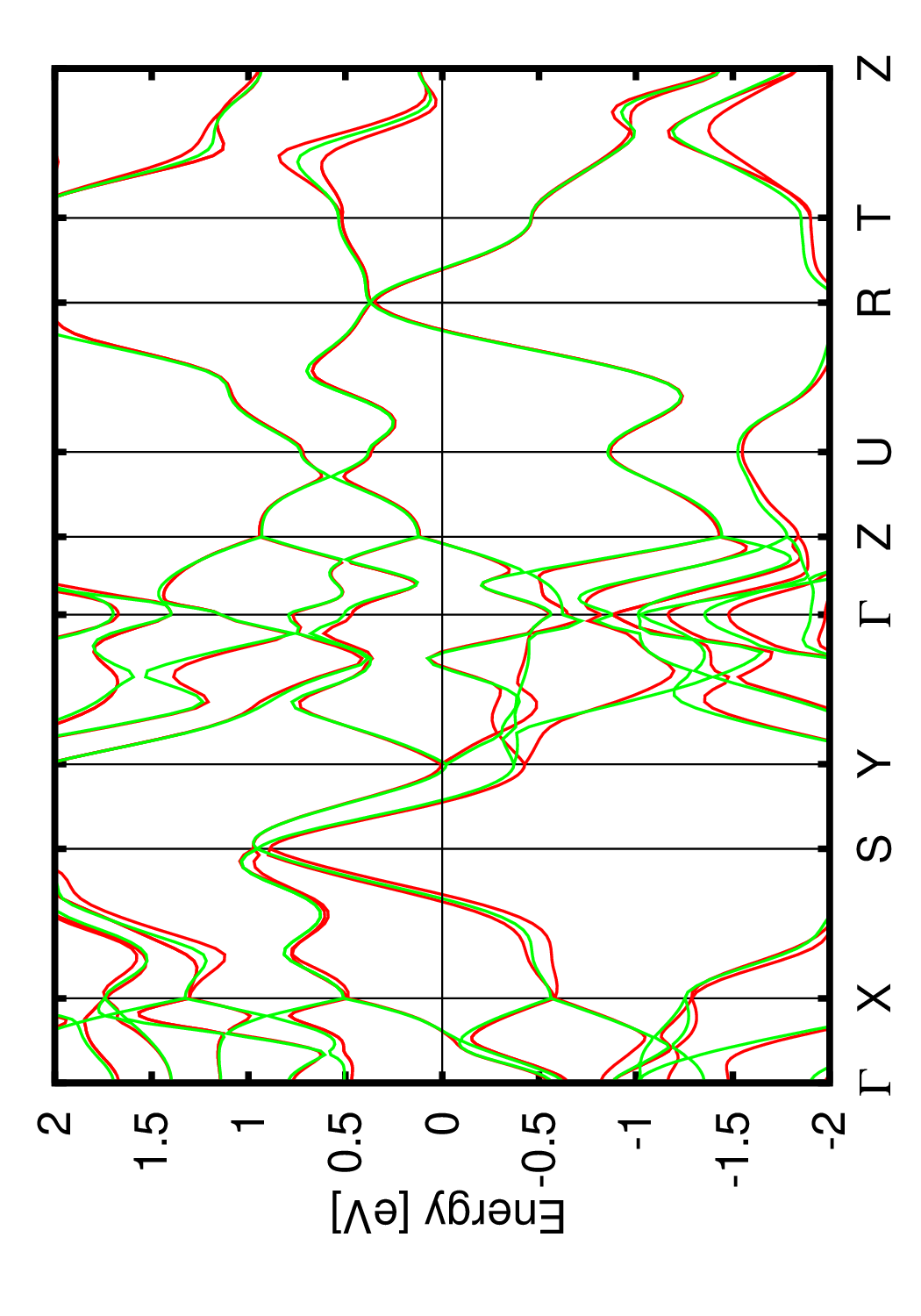}
	\caption{Band structure of the WP along the high-symmetry path of the orthorhombic BZ without (green lines) and with SOC (red lines). We plot in the energy range from -9 eV to +9 eV (top panel) and from -2 eV to +2 eV (bottom panel).  The Fermi level is set at the zero energy.
	}
	\label{Band1}
\end{figure}

\begin{figure}
	\centering
	\includegraphics[width=6cm,height=9cm,angle=270]{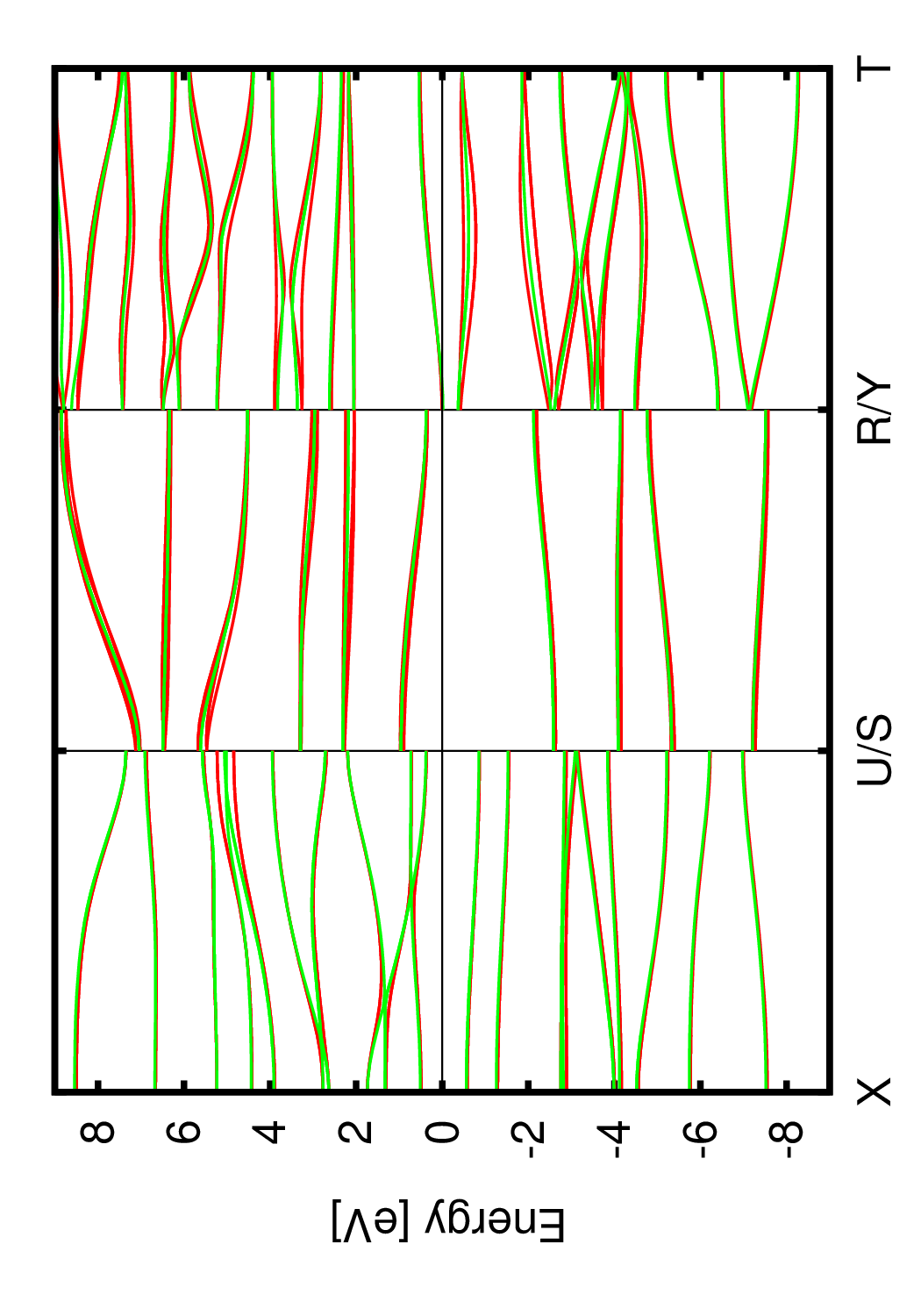}
	\includegraphics[width=6cm,height=9cm,angle=270]{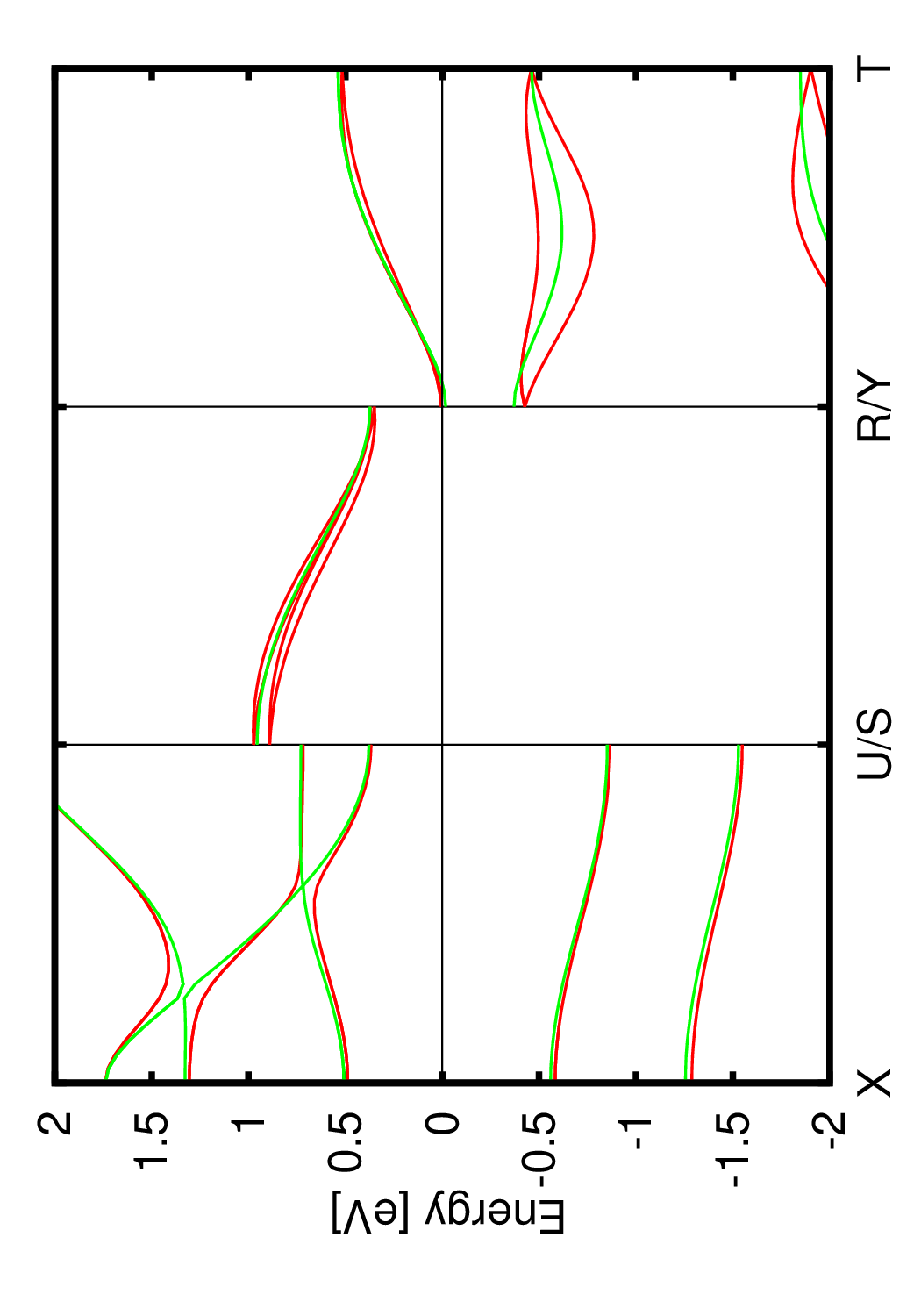}
	\caption{Band structure of the WP along the remaining high-symmetry lines of the orthorhombic BZ without (green lines) and with SOC (red lines). We plot in the energy range from -9 eV to +9 eV (top panel) and from -2 eV to +2 eV (bottom panel). The Fermi level is set at the zero energy.
	}
	\label{Band2}
\end{figure}

In Figs.~\ref{Band1}-~\ref{Band2} we show the band spectrum derived from first principle calculations, in the cases where the spin-orbit coupling in included or not in the calculations. The bandwidth is around 16 eV and, as already pointed out for the DOS, there are flat bands near the Fermi level, due to the W degrees of freedom, while the wider bands located far from the Fermi level are mainly due to the phosphorus atoms. 
 
The relevant features of the band structure without SOC may be ascribed to the symmetries of the underlined Hamiltonian that may describe the system. Since the system is symmetric under the unitary inversion operator $\hat{P}$ and upon the action of the time-reversal operator $\hat{T}$, the antiunitary inversion-time reversal $\hat{P}\hat{T}$ operator gives rise to bands that are two-fold degenerate at any k-vector of the BZ. This is the Kramers degeneracy and it is protected under the action of SOC. The system possess also nonsymmorphic symmetries,~\cite{Niu16} which
involve half translation of a Bravais lattice, causing an additional degeneracy along specific lines of the BZ. This additional degeneracy, which can be two-fold in case of one nonsymmorphic symmetry or four-fold in case of two nonsymmorphic symmetries, is active only along some high-symmetry lines because the corresponding operators are momentum-dependent.~\cite{Niu16}
In particular, along the SR line the band structure carries an overall eight-fold degeneracy due to the presence of two nonsymmorphic symmetries. Such multifold degeneracy is also expected to effectively reduce the probability for the bands to cross the Fermi level and thus they can induce some consequences on the dimensionality of the Fermi surface, as we will consider in more details in the next sections.

The interplay between the SOC interaction and the inter-orbital degrees of freedom allows for a selective removal along some lines of the BZ of the band degeneracy ascribed to the crystal symmetries. Specifically, the SOC partially removes the degeneracy of the bands along the XS, TZ, YT and SR high-symmetry lines, bringing them from four-fold to two-fold degeneracy along the first three and from eight-fold to two-fold degeneracy along the last one.
Instead, at the point R and S, the SOC partially removes the degeneracy bringing it from  eight- to four-fold. The splitting due to the SOC along the SR line is relatively small and it does not change the dimensionality of the Fermi surface. We would like to point out that the spin-orbit coupling constants $\lambda$ estimated from the real space Hamiltonian for the W and P atoms are $\lambda_{W}$=228~meV for the $5d$-orbitals of the W atoms and $\lambda_{P}$=55~meV for the $3p$-orbitals of the P atoms, respectively.
As a consequence the SOC splitting strongly affects the bands near the Fermi level because these bands mainly have a W character, and this effect turns out to be relevant for the transport properties. Although the SOC splitting is larger in WP compound than in CrAs and MnP systems,~\cite{Autieri17,Cuono19B}  the selective removal of the band degeneracy due to the crystal symmetries is also present in these latter materials.

Due to the eight-fold degeneracy and the presence of the two nonsymmorphic symmetries, each minimal model should contain at least four bands, which are equivalent to two bands per formula unit. As a consequence, the system will be always metallic if the time reversal symmetry protection is valid and the sum of $p$ and $d$ electrons per formula unit is odd, like in WP and CrAs. 

Because of the symmetries of the system, we will have semi-Dirac-like energy-momentum dispersions near the points X, Y and Z of the BZ. In particular, as already mentioned above, the semi-Dirac point at Y has been proposed to produce the linear magnetoresistence in CrAs and CrP.~\cite{Niu16,Niu19}
The band structure in Figs.~\ref{Band1}-~\ref{Band2} shows a semi-Dirac point at 0.02 eV below the Fermi level suggesting that linear dependent magnetoresistance could be also found in the WP compound.

We conclude this section noticing that the results here presented for the band spectrum and DOS share robust qualitative features with those of the CrAs~\cite{Autieri17} and MnP compounds.~\cite{Continenza01}

\section{Nonsymmorphic symmetries along the SR line}

In this section we demonstrate that the presence of nonsymmorphic symmetries brings to additional degeneracies along specific lines of the BZ in the case of vanishing SOC interaction, and is at the origin of the eight-fold degeneracy along the SR path the BZ.
We start by considering a tight-binding description of the WP electronic structure, whose matrix elements have been set according to the outcome of the DFT calculations. In particular, we restrict to the representative subspace of two $d$-orbitals (i.e. $d_{xy}$ and $d_{yz}$) and consider only non-vanishing projected W-W hopping amplitudes. When the spin degeneracy is considered, one can reduce the analysis by considering separately each spin channel, thus remaining with two $8\times8$ blocks in the Hamiltonian matrix.
We denote by $t_{\alpha_i,\alpha_j}^{lmn}$ the hopping amplitudes between the sites $\alpha_i$ and $\alpha_j$ (where $i,j=A,B,C,D$ as labelled in Fig.~\ref{Structure}) along the direction l$\mathbf{x}$ + m$\mathbf{y}$ + n$\mathbf{z}$. 
The Hamiltonian for each spin-channel is then written as follows:

\begin{equation}\label{HAM}
H(k_{x},k_{y},k_{z})=
\begin{pmatrix} 
H_{\alpha\alpha}(k_{x},k_{y},k_{z}) & H_{\alpha\beta}(k_{x}) \\
H_{\beta\alpha}(k_{x}) & H_{\beta\beta}(k_{x},k_{y},k_{z}) \\
\end{pmatrix},
\end{equation}

\noindent where $\alpha$ and $\beta$ label the $d_{xy}$ and the $d_{yz}$ orbital, and $H_{\beta\alpha}$ is the conjugate transpose of $H_{\alpha\beta}$.
\noindent The diagonal term of the matrix $H_{\alpha\alpha}$ is expressed as:

\begin{equation}
\label{haa}
\begin{split}
H_{\alpha\alpha}(k_{x},k_{y},k_{z})  = & t_{1}(k_{x},k_{y},k_{z}) s_{0} \otimes \tau_{0} + t_{2}(k_{x}) s_{0} \otimes \tau_{x} + \\ 
& + t_{3}(k_{x},k_{y},k_{z}) s_{x} \otimes \tau_{0} + \\
& + t_{4}(k_{x},k_{y},k_{z}) s_{y} \otimes \tau_{0} + \\
& + t_{5}(k_{x},k_{y},k_{z}) s_{x} \otimes \tau_{z} + \\
& + t_{6}(k_{x},k_{y},k_{z}) s_{y} \otimes \tau_{z} + \\
& + t_{7}(k_{x}) s_{z} \otimes \tau_{y}
\end{split}
\end{equation}

\noindent where $s_{i}$, $\tau_{i}$, $\sigma_{i}$ represent the Pauli matrices for $i=x,y,z$, and the unit matrices for $i=0$. $s_{i}$ act in the subspace of spins, $\tau_{i}$ in the subspace of orbitals, $\sigma_{i}$ and $\gamma_{i}$ in the subspace set by the atomic states in the unit cell. An analogous expression to Eq. \eqref{haa} defines also $H_{\beta\beta}$.\\
The hopping parameters have the following expressions:

\begin{equation}
\begin{split}
t_{1}(k_{x},k_{y},k_{z}) & = \varepsilon_{0} + 2\sum_{n=1,2,3}t^{n00}_{AA\alpha\alpha}\cos{(nk_x)} + \\
& + 2\sum_{n=1,2,3}t^{0n0}_{AA\alpha\alpha}\cos{(nk_y)} + \\
& + 2\sum_{n=1,2,3}t^{00n}_{AA\alpha\alpha}\cos{(nk_z)}  \\
t_{2}(k_{x}) & = t^{100}_{AB\alpha\alpha} (1 + cos(k_{x})) \\
t_{3}(k_{x},k_{y},k_{z}) & =\frac{1}{2}(t^{001}_{AC\alpha\alpha} + \\
& + t^{00\bar{1}}_{AC\alpha\alpha}cos(k_{x}) + t^{001}_{AC\alpha\alpha}cos(k_{y}) + \\
& + t^{00\bar{1}}_{AC\alpha\alpha}cos(k_{x}+k_{y}) + t^{00\bar{1}}_{AC\alpha\alpha}cos(k_{z}) + \\
& + t^{00\bar{1}}_{AC\alpha\alpha}cos(k_{y}+k_{z}) + t^{001}_{AC\alpha\alpha}cos(k_{x}+k_{z}) + \\
& + t^{001}_{AC\alpha\alpha}cos(k_{x}+ k_{y}+ k_{z})) \\
t_{4}(k_{x},k_{y},k_{z}) & = \frac{1}{2}(t^{00\bar{1}}_{AC\alpha\alpha}sin(k_{x}) + \\
& + t^{001}_{AC\alpha\alpha}sin(k_{y}) + t^{00\bar{1}}_{AC\alpha\alpha}sin(k_{x}+k_{y}) + \\
& + t^{00\bar{1}}_{AC\alpha\alpha}sin(k_{z}) + t^{00\bar{1}}_{AC\alpha\alpha}sin(k_{y}+k_{z}) + \\
& + t^{001}_{AC\alpha\alpha}sin(k_{x}+k_{z}) + \\
& + t^{001}_{AC\alpha\alpha}sin(k_{x}+k_{y}+k_{z})) \\
t_{5}(k_{x},k_{y},k_{z}) &  = \frac{1}{2}(t^{001}_{AC\alpha\alpha} - t^{00\bar{1}}_{AC\alpha\alpha}cos(k_{x}) + \\
& + t^{001}_{AC\alpha\alpha}cos(k_{y}) - t^{00\bar{1}}_{AC\alpha\alpha}cos(k_{x}+k_{y}) + \\
& + t^{00\bar{1}}_{AC\alpha\alpha}cos(k_{z}) + t^{00\bar{1}}_{AC\alpha\alpha}cos(k_{y}+k_{z}) + \\
& - t^{001}_{AC\alpha\alpha}cos(k_{x}+k_{z}) + \\
& - t^{001}_{AC\alpha\alpha}cos(k_{x}+k_{y}+k_{z})) \\
t_{6}(k_{x},k_{y},k_{z}) & = \frac{1}{2}(-t^{00\bar{1}}_{AC\alpha\alpha}sin(k_{x}) + \\
& + t^{001}_{AC\alpha\alpha}sin(k_{y}) - t^{00\bar{1}}_{AC\alpha\alpha}sin(k_{x}+k_{y}) + \\
& + t^{00\bar{1}}_{AC\alpha\alpha}sin(k_{z}) + t^{00\bar{1}}_{AC\alpha\alpha}sin(k_{y}+k_{z}) + \\
& - t^{001}_{AC\alpha\alpha}sin(k_{x}+k_{z}) + \\
& - t^{001}_{AC\alpha\alpha}sin(k_{x}+k_{y}+k_{z})) \\
\end{split}
\end{equation}
\begin{equation}
\begin{split}
t_{7}(k_{x}) &  = - t^{100}_{AB\alpha\alpha}sin(k_{x})
\end{split}
\end{equation}

\noindent The off-diagonal term $H_{\alpha\beta}$ is written as:

\begin{equation}
\begin{split}
H_{\alpha\beta}(k_{x}) & =
\begin{pmatrix} 
t^{000}_{AA\alpha\beta} & t^{100}_{AB\alpha\beta} & 0 & 0 \\
t^{100}_{AB\beta\alpha} & t^{000}_{AA\alpha\beta} & 0 & 0 \\
0 & 0 & t^{000}_{AA\alpha\beta} & t^{100}_{AB\alpha\beta}\\
0 & 0 & t^{100}_{AB\beta\alpha} & t^{000}_{AA\alpha\beta}\\
\end{pmatrix} + \\
& +
e^{i k_{x}}
\begin{pmatrix} 
0 & t^{100}_{AB\beta\alpha} & 0 & 0 \\
0 & 0 & 0 & 0 \\
0 & 0 & 0 & 0\\
0 & 0 & t^{100}_{AB\alpha\beta} & 0\\
\end{pmatrix}+ \\
& +
e^{-i k_{x}}
\begin{pmatrix} 
0 & 0 & 0 & 0 \\
t^{100}_{AB\alpha\beta} & 0 & 0 & 0 \\
0 & 0 & 0 & t^{100}_{AB\beta\alpha}\\
0 & 0 & 0 & 0\\
\end{pmatrix}.
\end{split}
\end{equation}

\noindent where $t^{100}_{AB\alpha\beta}\not=t^{100}_{AB\beta\alpha}$. Note that all the hoppings are written in units of $a$, $b$ and $c$.\\
We will now focus on the tight-binding Hamiltonian along the SR line $H(k_z)$, where we keep fixed ($k_{x}$,$k_{y}$)=($\pi$,$\pi$). We notice that, along that high symmetry line, the $t_{2}(k_{x})$,...,$t_{7}(k_{x})$ hybridization terms in the intra-orbital subspace do vanish. This circumstance reduces $H(k_z)$ in a non block-diagonal form, where only the intra-orbital sectors $H_{\alpha \alpha}(k_z)$ and $H_{\beta \beta}(k_z)$ are diagonal.\\
	
\noindent When we take both spin-channels, we can then consider the two unitary operators:

\begin{equation}
\label{a}
\hat{Q}=s_{0} \otimes \tau_{0} \otimes \sigma_{y} \otimes \gamma_{y}
\end{equation} 

\begin{equation}
\label{b}
\hat{U}(k_{z})=\begin{pmatrix} 
0 & 1 \\
e^{ik_{z}} & 0 
\end{pmatrix} \otimes \tau_{0} \otimes \sigma_{z} \otimes \gamma_{y}\ ,
\end{equation}

\noindent where $\gamma_{i}$ represent the Pauli matrices for $i=x,y,z$, and the unit matrices for $i=0$. The operators $\hat{Q}$ and $\hat{U}(k_{z})$ commute with the Hamiltonian $H(k_z)$ along the SR path:

\begin{equation}
\label{c}
\comm{\hat{H}}{\hat{Q}}=0 , \quad  at ~  (k_{x},k_{y})=(\pi,\pi)
\end{equation}

\begin{equation}
\label{d}
\comm{\hat{H}}{\hat{U}(k_{z})}=0, \quad at ~ (k_{x},k_{y})=(\pi,\pi) \ .
\end{equation}

\noindent Moreover, they anticommute with each other:

\begin{equation}
\label{e}
\acomm{\hat{Q}}{\hat{U}(k_{z})}=0.
\end{equation}

\noindent In particular, $\hat{Q}$ acts by exchanging the coordinates of A and D atoms within the unit cell; $\hat{U}(k_{z})$ is instead a nonsymmorphic operator involving half translation of the reciprocal lattice and a transformation on the unit cell configurations, which satisfies the following relation:

\begin{equation}
\label{f}
\hat{U}^{2}(k_{z}) = e^{ik_{z}} \mathbb{I}\ ,
\end{equation}

\noindent whose eigenvalues are $\pm e^{ik_{z}/2}$, and where $\mathbb{I}$ is the identity matrix.~\cite{Zhao16}

\noindent Along the SR line, $H(k_z)$ also commutes with other two unitary operators:

\begin{equation}
\label{g}
\hat{V}=s_{0} \otimes \tau_{0} \otimes \sigma_{z} \otimes \gamma_{y}
\end{equation} 

\begin{equation}
\label{h}
\hat{W}(k_{z})=\begin{pmatrix} 
0 & 1 \\
e^{ik_{z}} & 0 
\end{pmatrix} \otimes \tau_{0} \otimes \sigma_{y} \otimes \gamma_{y}
\end{equation} 

\begin{equation}
\label{i}
\comm{\hat{H}}{\hat{V}}=0 , \quad at ~ (k_{x},k_{y})=(\pi,\pi)
\end{equation}

\begin{equation}
\label{l}
\comm{\hat{H}}{\hat{W}(k_{z})}=0, \quad {\text{at}} ~ (k_{x},k_{y})=(\pi,\pi)
\end{equation}

\noindent Moreover, $\hat{V}$ and $\hat{W}(k_{z})$ anticommute with each other:
\begin{equation}
\label{m}
\acomm{\hat{V}}{\hat{W}(k_{z})}=0,
\end{equation}

\noindent We notice that $\hat{V}$ acts by exchanging the coordinates of B and C atoms within the unit cell, while $\hat{W}(k_{z})$, apart from acting on the intra-unit cell orbital degrees of freedom, is a nonsymmorphic operator involving half translation of the reciprocal lattice.\\
Furthermore, we have that the operators $\hat{Q}$, $\hat{V}, \hat{W}(k_{z})$ and $\hat{U}(k_{z})$ fulfill the following algebra:

\begin{equation}
\label{n}
\acomm{\hat{Q}}{\hat{V}}=0,
\end{equation}

\begin{equation}
\label{o}
\comm{\hat{Q}}{\hat{W}(k_{z})}=0\ .
\end{equation}

\begin{equation}
\label{p}
\comm{\hat{U}(k_{z})}{\hat{V}}=0,
\end{equation}

\begin{equation}
\label{q}
\comm{\hat{U}(k_{z})}{\hat{W}(k_{z})}=0,
\end{equation}

\noindent Since $\hat{U}$ and $\hat{V}$ commute between each other and with the Hamiltonian, it is possible to choose a common eigenstate $\psi_{uv}$, labeled by the corresponding $u$ and $v$ eigenvalues of $\hat{U}$ and $\hat{V}$, associated to the same energy $E(k_z)$:

\begin{equation}
\label{r}
\hat{U}\psi_{uv}=u\psi_{uv} 
   \quad\text{}\quad 
\hat{V}\psi_{uv}=v\psi_{uv}\ .
\end{equation}
\begin{equation}
\label{hz}
H(k_z)\psi_{uv}=E(k_z)\psi_{uv} 
\end{equation}

\noindent Then, by using the commutation and anticommutation relations between the above operators (see the Eqs.~\eqref{e}, ~\eqref{m} and ~\eqref{n}~--~\eqref{q}), one can have:

\begin{equation}
\label{s}
\hat{U}(\hat{Q}\psi_{uv})=-u(\hat{Q}\psi_{uv}) 
   \quad\text{}\quad 
\hat{V}(\hat{Q}\psi_{uv})=-v(\hat{Q}\psi_{uv}),
\end{equation}

\begin{equation}
\label{t}
\hat{U}(\hat{W}\psi_{uv})=u(\hat{W}\psi_{uv})
\quad\text{}\quad 
\hat{V}(\hat{W}\psi_{uv})=-v(\hat{W}\psi_{uv}),
\end{equation}

\begin{equation}
\label{u}
\hat{U}(\hat{Q}\hat{W}\psi_{uv})=-u(\hat{Q}\hat{W}\psi_{uv})
\quad\text{}\quad 
\hat{V}(\hat{Q}\hat{W}\psi_{uv})=v(\hat{Q}\hat{W}\psi_{uv}),
\end{equation}

\noindent Eqs.~\eqref{r}~--~\eqref{u} show that it is possible to construct a subspace associated to the same energy eigenvalue, spanned by the four orthogonal basis states: $\psi_{uv}$, $\hat{Q}\psi_{uv}$, $\hat{W}\psi_{uv}$ and $\hat{Q}\hat{W}\psi_{uv}$. The dimensionality of this subspace equals the four-fold degeneracy which is obtained along the SR line, for each of the spin channels. We finally point ot that the inclusion of the SOC interaction in the tight-binding Hamiltonian partially removes the degeneracy, bringing it from eight-fold to two-fold. This is because the total Hamiltonian is no longer invariant upon $\hat{Q}$, $\hat{U}$, $\hat{V}$ and $\hat{W}$. On the contrary, the SOC interaction does not remove the time-inversion degeneracy associated to the combination of inversion and time reversal symmetry. The results of the presented analysis allow us to conclude that the four-fold degeneracy for each spin channel is due to the existence of two couples of anticommuting symmetry operators, two of them being nonsymmorphic, which satisfy the algebra given by Eqs.~\eqref{e}, ~\eqref{m} and ~\eqref{n}~--~\eqref{q}.

\section{Effect of the nonsymmorphic symmetries on the Fermi surface}

In this section we investigate the evolution of the Fermi surfaces (FS) as a function of the electron filling. 
In the first subsection, we will discuss some general consideration while in the second we will report the DFT results.
In the second subsection, we will start by considering the case of vanishing SOC, while at the end of this section the effects of the SOC term will be discussed.\\

\begin{figure*}[ht!]
\centering
\includegraphics[width=\columnwidth, height=7cm, angle=0]{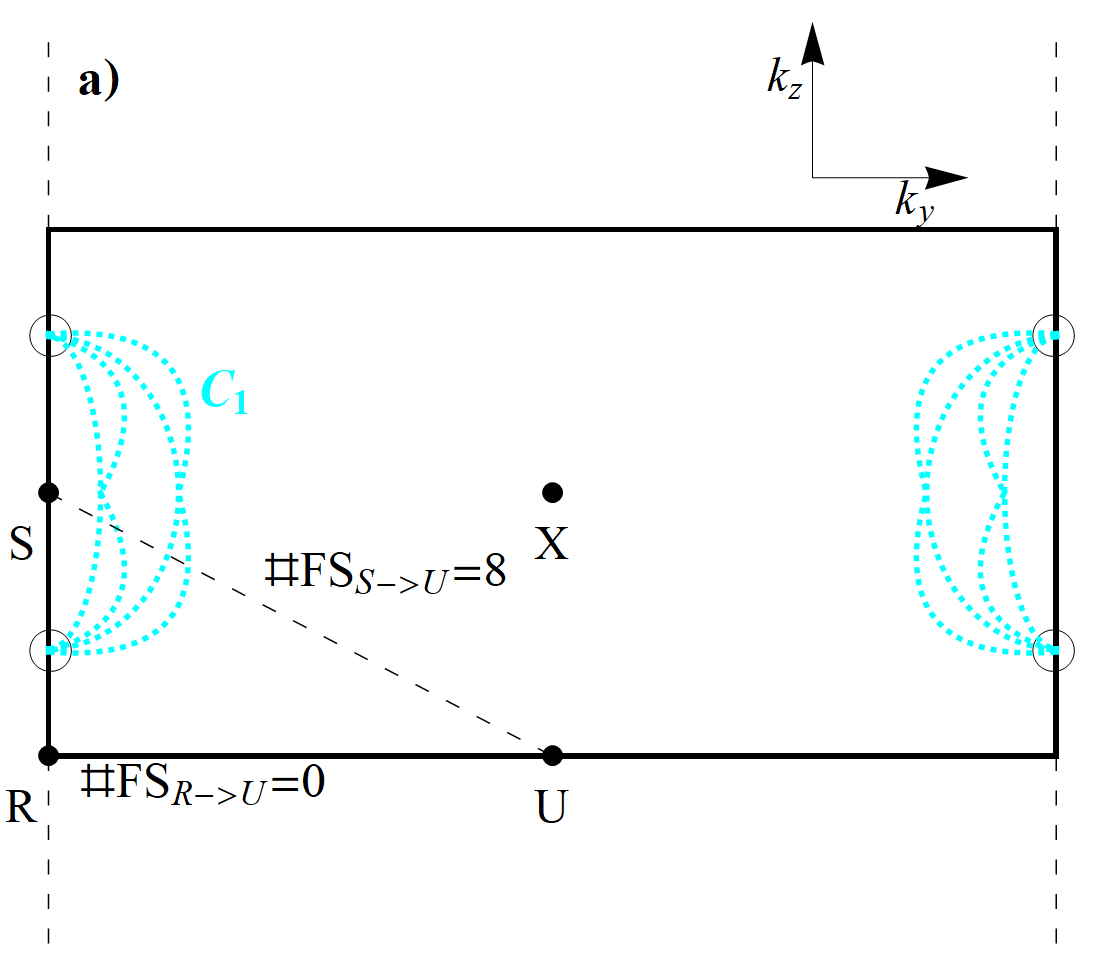}
\includegraphics[width=\columnwidth, height=7cm, angle=0]{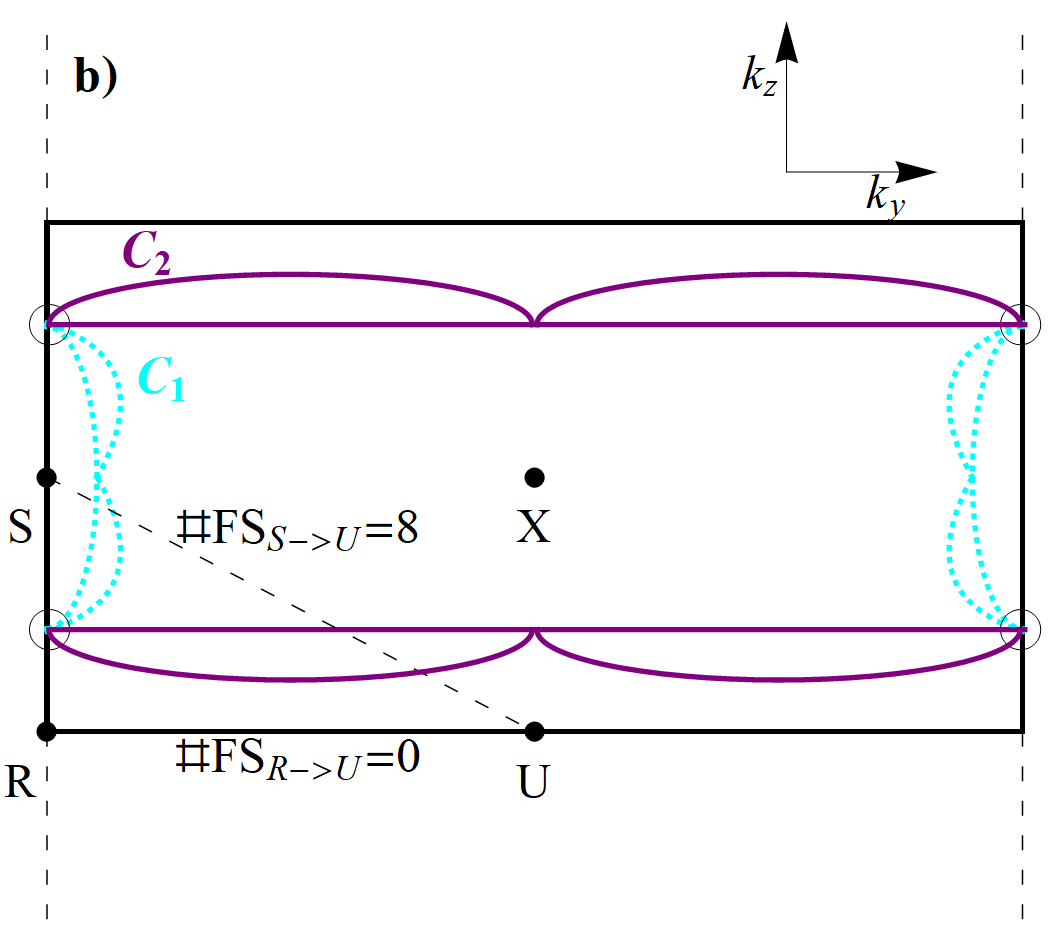}
\includegraphics[width=\columnwidth, height=7cm, angle=0]{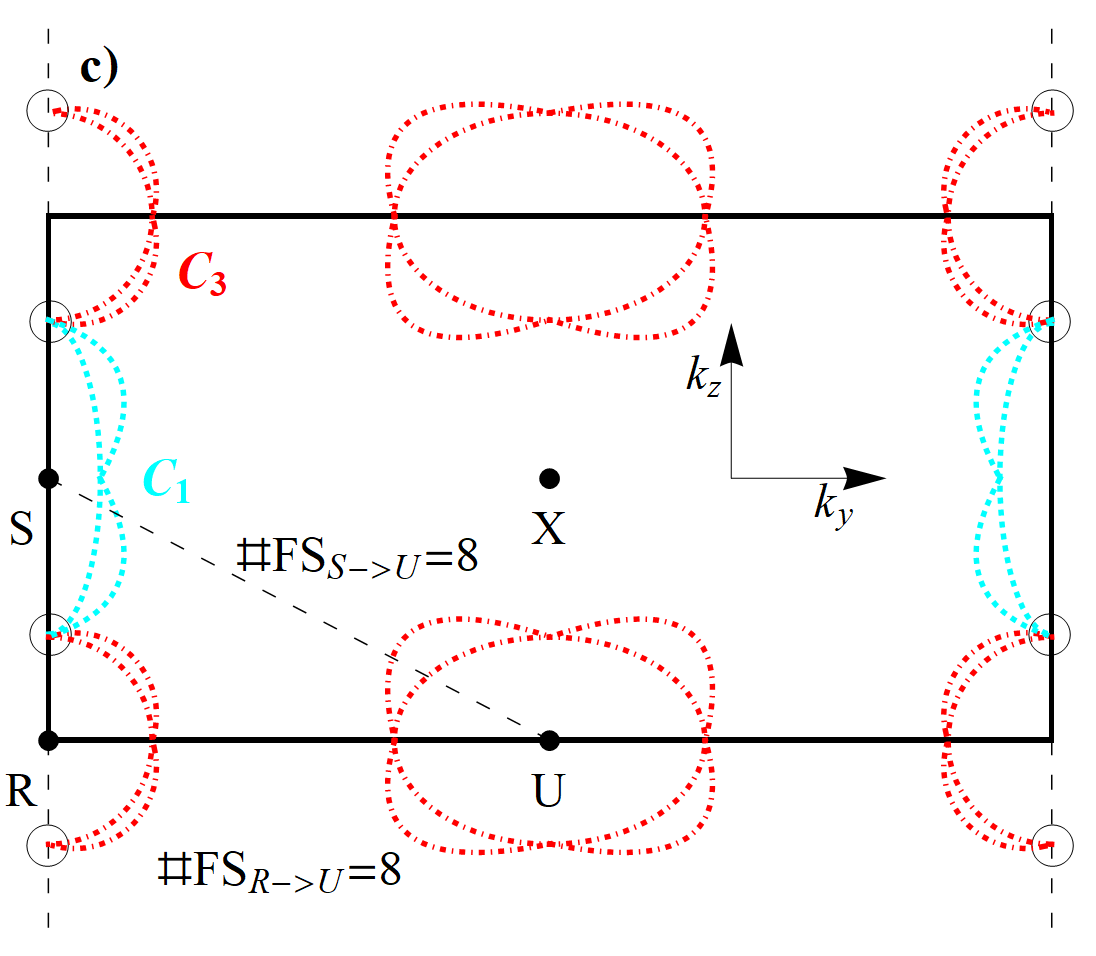}
\includegraphics[width=\columnwidth, height=7cm, angle=0]{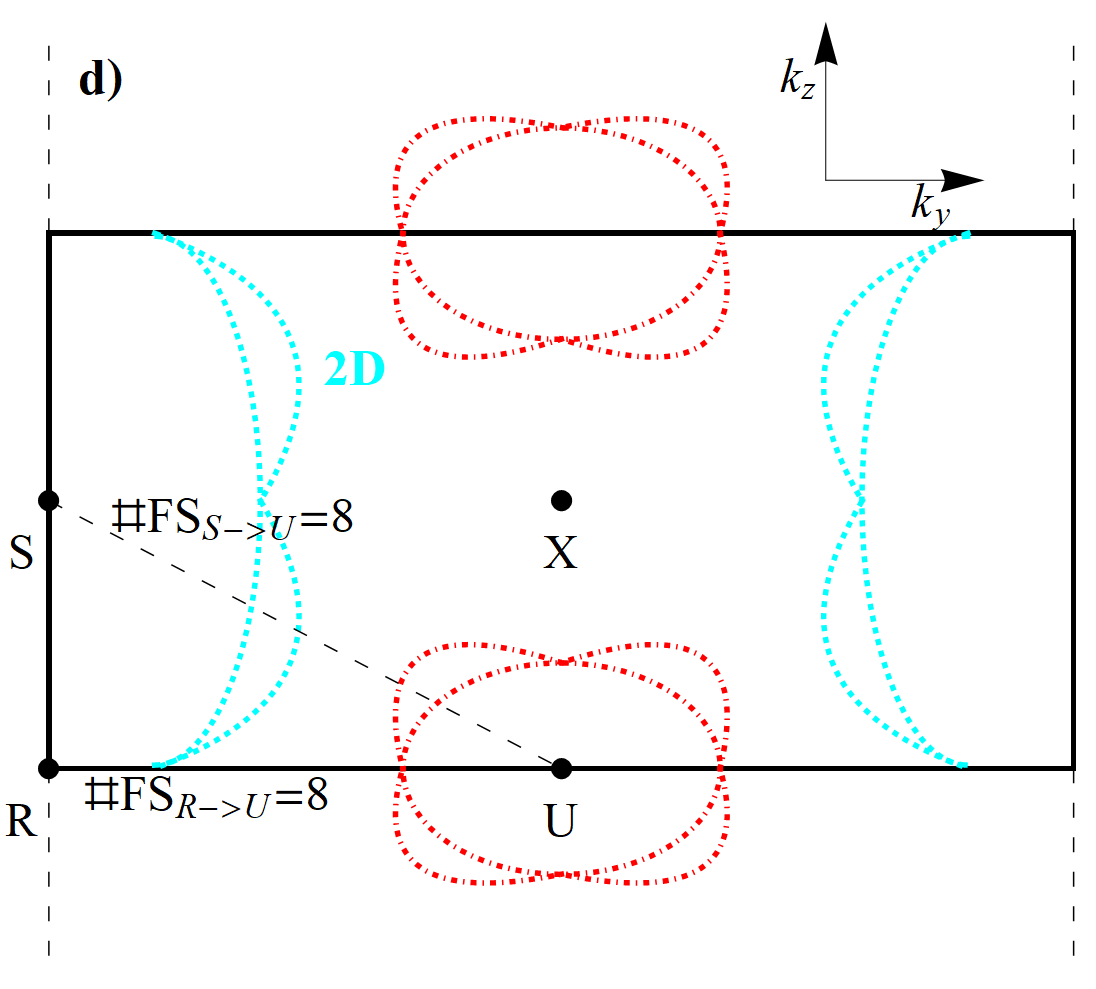}
\caption{(k$_y$,k$_z$) plane side view of the connections obtained in the MnP compound. a) Schematic Fermi surface for n$_2\ll$n<n$_1$.
b) Schematic Fermi surface for n$_2$<n$\ll$n$_1$. 
c) Schematic Fermi surface for n$_3$<n<n$_2$.
d) Schematic Fermi surface for n<n$_3$. The Fermi points at ($\pm\pi,\pm\pi,\pm k_F$) are represented by the open circles in the a), b) and c) panels.
The critical fillings n$_1$, n$_2$ and n$_3$ are defined in Fig. \ref{BSMnP}.}
\label{FS_cartoon}
\end{figure*}

The FS of the WP are deduced at ambient pressure whereas those of the CrAs and MnP are derived at the critical  pressures where the magnetism is suppressed and the superconductivity comes in. In the case of CrAs, we use the results from the literature~\cite{Shen16}.
Instead, for the MnP the experimentally measured pressure amplitude that suppresses the magnetism is 7 GPa,~\cite{Gercsi10} while from our {\it ab-initio} calculations we find that the magnetization goes to zero with a metamagnetic transition when the volume is uniformly reduced by 5.8~\%.
Therefore, we apply the theoretically predicted volume, which is reduced of 5.8~\% with respect to the ambient pressure, for proceeding further about the determination of the Fermi surface. For completeness, we mention that the metamagnetic transition can be accounted by considering the peaks of the DOS close to Fermi level.~\cite{Gercsi10,Autieri12}

\subsection{General considerations}

We first consider the case of Fermi points along the SR line. 
If the bands along the SR direction cross the Fermi level, we have eight Fermi points ($\pm\pi,\pm\pi,\pm k_F$) due to the inversion and mirror symmetry in this class of compounds. The Fermi points ($\pm\pi,\pm\pi,\pm k_F$) are eight-fold degenerate, four-fold because of the intracell orbital degrees of freedom, which are connected by two nonsymmorphic symmetries, and two-fold because of the inversion-time reversal symmetry.
Since the electronic structure along the SR line does not have a semimetallic behavior, due to the time-inversion degeneracy in the whole BZ, in any plane that includes the SR line there are four Fermi lines emerging from the Fermi points which can evolve in various ways by ending into the other Fermi points.
Many combinations are possible, however, an even number of Fermi points should be connected because of inversion and mirror symmetries.

At this stage we would like to discuss the connection between the degeneracy of electronic states along the high symmetry lines and the dimensionality and topology of the multiple Fermi surface sheets.  
Let us start with a basic observation. The fact of having an eightfold degenerate Fermi point along the $SR$ line implies that there will be four separate Fermi  lines (i.e. two times four due to the twofold degeneracy arising from the time-inversion symmetry in the whole BZ) around it in any given plane containing the Fermi point itself. These lines can start from one of the Fermi points along the $SR$ line and then end into another one or in the same passing through the BZ or extending in a limited region close to the high symmetry line. Since different lines can have inequivalent connecting properties, the resulting topological aspect of the closed Fermi loops is highly nontrivial and its evolution is tied to the multifold character crossing points within the Brilluoin zone.

As example of this class of compounds, we examine the case of the MnP that shows many different Fermi surface topologies close to the nominal filling. We show (k$_y$,k$_z$) side view of the FS sketch since it is the only one interesting in this case.  
In order to be more illustrative, we have deviced different topological configurations of the connecting lines between the Fermi points on the $SR$ high symmetry line (see Figs.~\ref{FS_cartoon} (a)-(c)). 
A distinct state is then represented by an evolved configuration with respect to the previous ones but without Fermi points along the $SR$ line. This is represented in Fig.~\ref{FS_cartoon} (d) where the connecting lines are between Fermi points belonging to the $RU$ cut. 
In Fig. \ref{FS_cartoon} (a) we have depicted a situation where all the lines connect the two Fermi points without intersecting the zone boundaries. This type of connecting line is indicated as $C_1$. Another possibility is that some of the lines connect the same Fermi point by crossing the whole BZ and are labeled as $C_2$ and others are $C_1$-like (see Fig. \ref{FS_cartoon}(b)). Finally, we report a relevant physical case for the MX studied compounds, corresponding to the presence of two Fermi pockets centered around the $S$ and $R$ points of the BZ (see Fig. \ref{FS_cartoon} (c)), respectively.

To move further in the discussion, we observe that the cases in Figs.~\ref{FS_cartoon}(a) and (b) can be connected if two of the $C_1$ lines get closer to the X point and cross with the mirror symmetric ones. On the other hand, the configuration in Fig.~\ref{FS_cartoon} (b) can evolve into that one in Fig. \ref{FS_cartoon} (c) if the $C_2$ lines cross the $RU$ cut. 
These modifications of the Fermi lines are not directly linked with the multifold degeneracy along the $SR$ line. On the contrary, we would like to focus on the electronic transition between the case with two Fermi pockets around $S$ and $R$ (Fig.~\ref{FS_cartoon}(c)) and that one with two open Fermi lines depicted in Fig. \ref{FS_cartoon}(d). Indeed, if the change of filling is moving the Fermi points from $S$ to $R$, then the Fermi pockets associated to the lines $C_3$ shrink and disappear while, due to the fourfold degeneracy along the $RU$ cut, the $C_1$ lines keep staying connected with the Fermi points now evolving along the $RU$ cut. Such electronic transition is constrained by the fact that there is an eightfold degeneracy along the $SR$ line while it is only fourfold along the $RU$ cut. We recall that the degeneracy of the $RU$ line is dictated by the presence of only one nonsymmorphic symmetry transformation instead of two as for the $SR$ cut.
The electronic changeover exhibits a simultaneous modification of the Fermi surface dimensionality and topology. Indeed, the $C_3$ Fermi pockets disappear (red dots) while the $C_1$ pockets become open creating the 2D FSs.

Another remark, concerning a soft connection of the multifold degeneracy and the topology of the Fermi surface, is that when the nonsymmorphic symmetries are absent, the hybridization between the bands create the bonding-antibonding splitting producing a large bandwidth, so that the bands have higher probability to cross the Fermi level because they have a significant overlap with the other bands in the BZ. On the other hand, in the presence of the two nonsymmorphic symmetries, the electronic states collapse onto a unique band and the smaller bandwidth reduces the probability for the bands to cross the Fermi level.

\begin{figure}[h]
\centering
\includegraphics[height=\columnwidth, angle=270]{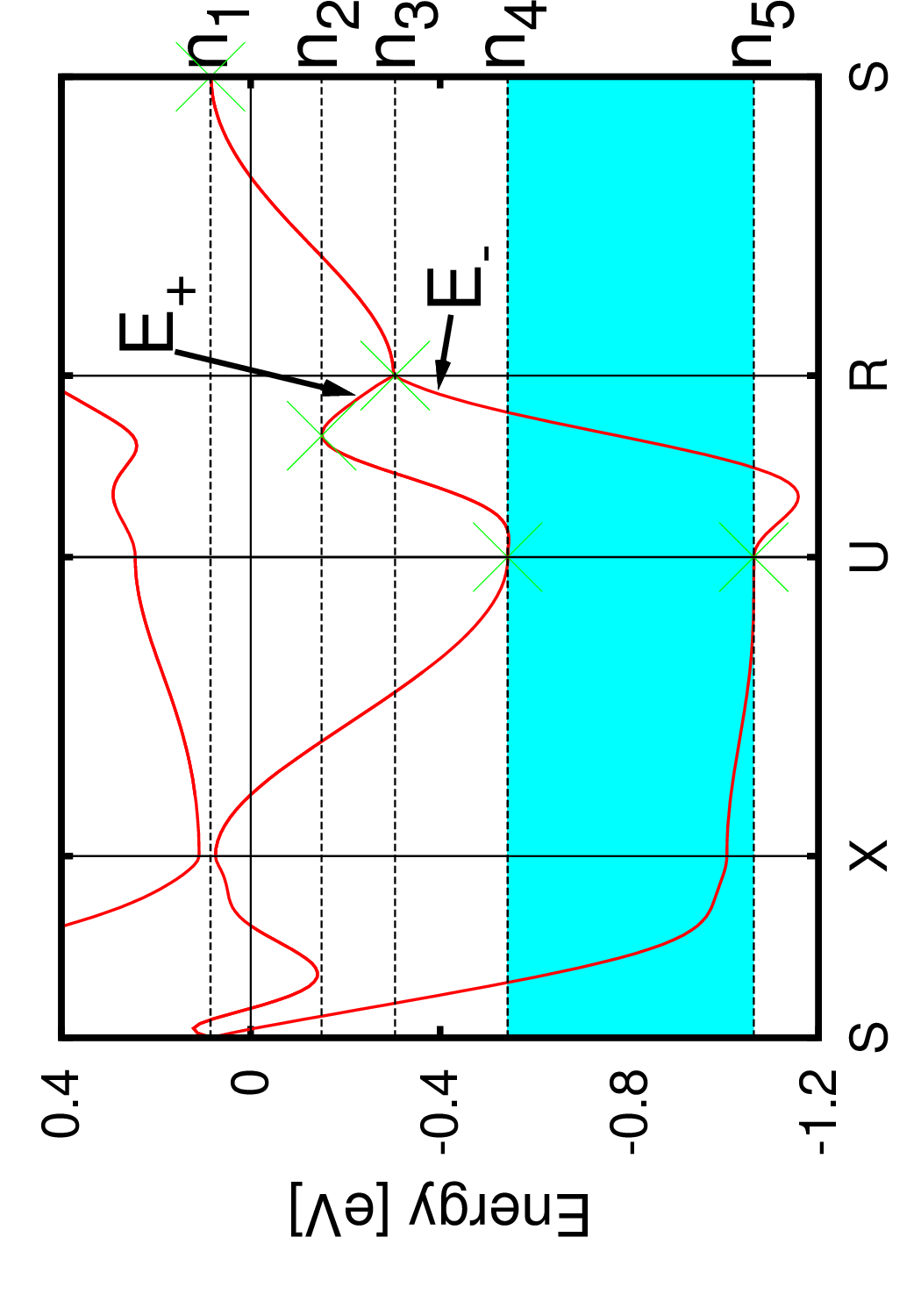}
\caption{Band structure of MnP along the S-X-U-R-S high-symmetry lines of the orthorhombic BZ without SOC. We plot in the energy range from -1.2 eV to +0.4 eV.  The topological region is shown in cyan. The Fermi level at the nominal filling is set at the zero energy.}
\label{BSMnP}
\end{figure}

Now, let us move to statements more specific about the MX class of compounds.
The band structure that is 8-fold degenerate at R by symmetry splits in two fourfold degenerate branches along RU as shown in Fig. \ref{BSMnP}.
When the band structure can be described with an effective single orbital, the Hamiltonian (\ref{HAM}) is suffice to describe the low energy electronic structure. From the Hamiltonian, we obtain that these fourfold degenerate branches have linear and opposite behaviour (E$_{+}$ and E$_{-}$ in Fig. \ref{BSMnP}), as we show more in detail in the Appendix D.
When the branch along SR is monotonic, the RU branch in the same energy range of the SR branch generates the four C$_3$ FSs while
the RU branch with opposite slope generates four open FS. If the branch along SR is weakly dispersive, the four open FSs are mostly 2D.
In summary, if the SR branch is monotonic, weakly dispersive and can be described by a single orbital as likely in this class of compounds, 
these systems show four 2D FSs for a certain filling.

\subsection{DFT results}

At this stage it is useful to follow the band structure of the MnP along the path S-X-U-R-S (see Fig. \ref{BSMnP}) and investigate the evolution of the Fermi surfaces as function of the filling of the system $n$ within DFT.
The nominal filling for the MnP is $n=10$ per formula unit, while the numerical values for the critical fillings, associated with the presence of Fermi points along the $SR$ cut are $n_1$=10.1 and $n_3$=9.3 per formula unit. The critical filling $n_2$ is related to the maximum of the band structure along the RU line. The critical fillings n$_4$=8.9 and n$_5$=8.2 are linked to the energy eigenvalues at the U point.
These critical fillings define regions where we can observe different Fermi surfaces topologies.\\

\begin{figure}[h]
\centering
\includegraphics[width=\columnwidth, angle=0]{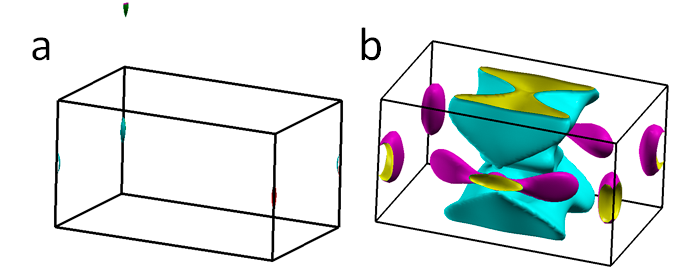}
\caption{Fermi surface of MnP in the first BZ without SOC with filling n$_2\ll$n<n$_1$.  
In panel a) we show two Fermi surfaces emerging from the Fermi points along the RS line, while in panel b) all the other Fermi surfaces.
For a better visualization we plot one FS in the BZ and another in the reciprocal unit cell.
}
\label{FSMnP1}
\end{figure}

\begin{figure}[h]
\centering
\includegraphics[width=\columnwidth, angle=0]{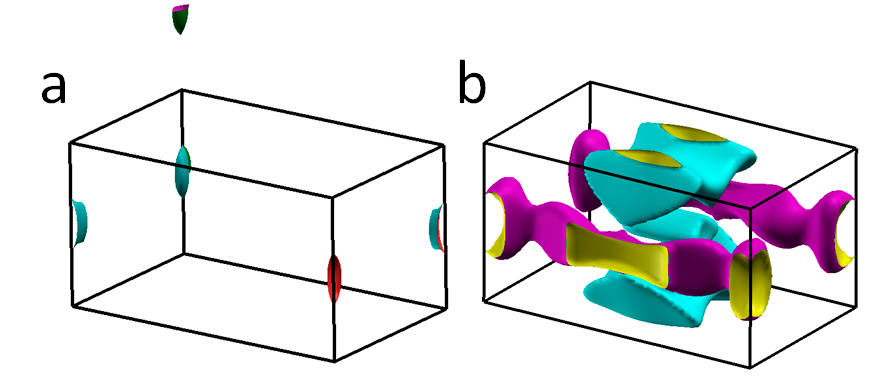}
\caption{Fermi surface of MnP in the first BZ without SOC with filling n$_2$<n$\ll$n$_1$. 
In panel a) we show two Fermi surfaces emerging from the Fermi points along the RS line, while in panel b) all the other Fermi surfaces.
For a better visualization we plot one FS in the BZ and another in the reciprocal unit cell.
}
\label{FSMnP2}
\end{figure}

\begin{figure}[h]
\centering
\includegraphics[width=\columnwidth, angle=0]{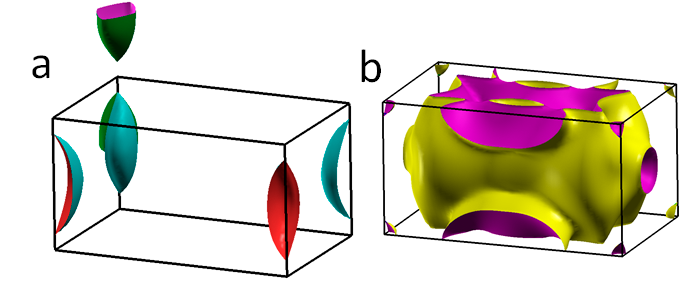}
\caption{Fermi surface of MnP in the first BZ without SOC with filling n$_3$<n<n$_2$.  
In panel a) we show two Fermi surfaces emerging from the Fermi points along the RS line, while in panel b) all the other Fermi surfaces.
For a better visualization we plot one FS in the BZ and another in the reciprocal unit cell.
}
\label{FSMnP3}
\end{figure}

When n is slightly larger than n$_1$, we have some Fermi pockets close to the S point but we do not have Fermi points along the RS line. 
When n=n$_1$ we have an eight-fold Fermi point at S.
When n$_2\ll$n<n$_1$ we have four connections of the kind C$_1$ as schematically shown in Fig. \ref{FS_cartoon}a). Two connections from the real band structure are shown in Fig.~\ref{FSMnP1}a) and two are shown in Fig.~\ref{FSMnP1}b).
When n$_2$<n$\ll$n$_1$ we have two connections of the kind C$_1$ in Figs.~\ref{FSMnP2}a) and two connection of the kind C$_2$ in Fig.~\ref{FSMnP2}b) while the simplified scheme is shown in Fig.~\ref{FS_cartoon}b).
When n$_3$<n<n$_2$ we have two connections of the kind C$_1$ in Figs.~\ref{FSMnP3}a) and two connections of the kind C$_3$ in Fig.~\ref{FSMnP3}b), the 
schematic FS is shown in Fig.~\ref{FS_cartoon}c).
Below n$_3$, the C$_1$ FSs evolve in 2D FSs. Between, n$_4$ and n$_5$ the system can sustain topological superconductivity.
In summary, two connections are always of the kind C$_1$, while the other two evolve from C$_1$ to C$_2$ and finally to C$_3$.

\begin{figure}[h]
\centering
\includegraphics[width=\columnwidth, angle=0]{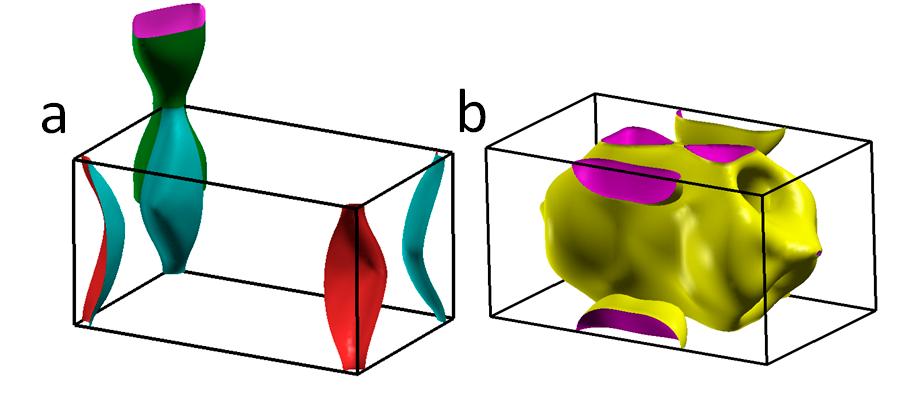}
\caption{Fermi surface of MnP in the first BZ without SOC with filling n$_4$<n<n$_3$.  
In panel a) we show two Fermi surfaces emerging from the Fermi points along the RS line, while in panel b) all the other Fermi surfaces.
For a better visualization we plot one FS in the BZ and another in the reciprocal unit cell.
}
\label{FSMnP4}
\end{figure}

\noindent When n=n$_3$ the Fermi points reach the R point. Reducing n and considering that the Fermi surfaces of the kind C$_1$ cannot disappear, the Fermi points can only move along k$_y$ forming the 2D Fermi surfaces. This implies that there will be open Fermi surfaces almost parallel to k$_z$.
Finally, when n<n$_3$ the two Fermi pockets of the kind C$_1$ evolve in two 2D-Fermi surfaces, while the two Fermi pockets of the kind C$_3$ disappear.
We thus obtain the 2D FS, as we show in Fig.~\ref{FSMnP4}a) and ~\ref{FSMnP4}b).

We now proceed with the explicit comparison of the Fermi surfaces (FS) of the normal phase of the CrAs, MnP and WP compounds. We point out that those materials are characterized by different fillings, which fall into the distinct regimes above considered. The number of pnictide-$p$ and metal-$d$ bands is 16 per formula unit. For CrAs and WP there are 9 electrons per formula unit, while in the case of MnP we have 10 electrons per formula unit. In the case of CrAs and WP, the bands along the SR line do not cross the Fermi level forming two-dimensional Fermi surface sheets centered around the SR line at ($k_{x}$,$k_{y}$)=($\pi$,$\pi$) and propagating along the $k_{z}$ direction.

\begin{figure}[h]
\centering
\includegraphics[width=\columnwidth, angle=0]{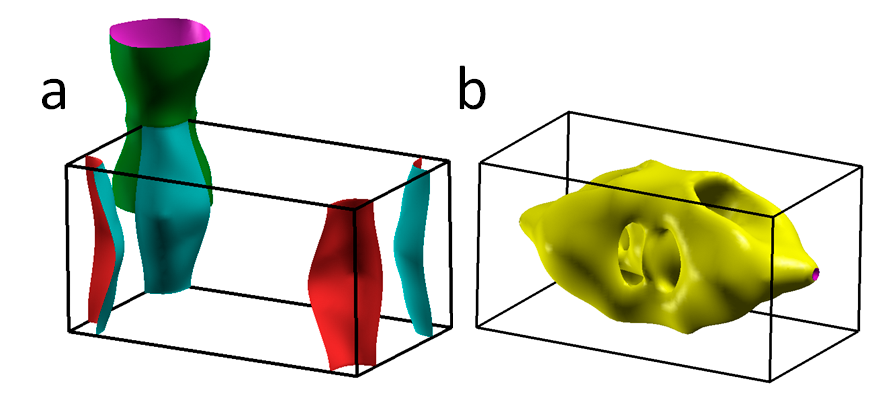}
\caption{Fermi surface of WP at ambient pressure in the first BZ without SOC. In panel a) we show the 2D sheets, while in panel b) the three-dimensional surface around the $\Gamma$ point. Since the two 2D FS are concentric, for a better visualization we plot one FS in the BZ and another in the reciprocal unit cell.
}
\label{FSWP}
\end{figure}

Let us discuss the 2D and 3D features of the FS of the WP plotted in Fig.~\ref{FSWP}a) and ~\ref{FSWP}b). The FS displays a three-dimensional surface around the $\Gamma$ point and two holelike 2D hourglass shaped sheets centred around the SR high-symmetry line, namely at ($k_{x}$,$k_{y}$)=($\pi$,$\pi$), and in the $ab$ plane.
We have to point out that the bands along the SR line are eight-fold degenerate, but when we move along the XS and the SY they splits in two four-fold bands.
One of this four-fold bands crosses the Fermi level, as we can see from the bands along the XS and SY lines plotted in Fig.~\ref{Band1}.

\begin{figure}[h]
\centering
\includegraphics[width=\columnwidth, angle=0]{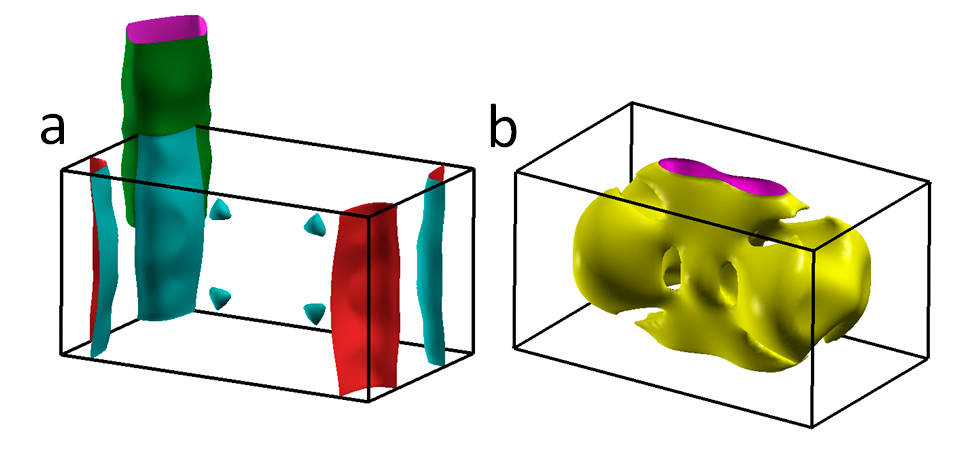}
\caption{Fermi surface of CrAs at P=0.94 GPa in the first BZ without SOC. In panel a) we show the 2D sheets, while in panel b) the three-dimensional surface around the $\Gamma$ point. Since the two 2D FS are concentric, for a better visualization we plot one FS in the BZ and another in the reciprocal unit cell.
}
\label{FSCrAs}
\end{figure}

\begin{figure}[h]
	\centering
	\includegraphics[width=\columnwidth, angle=0]{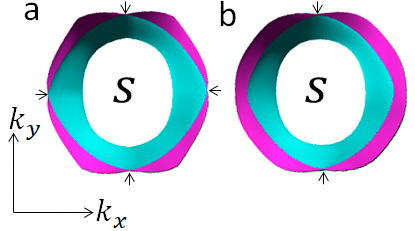}
	\caption{Top view of the 2D Fermi surface of WP centred at S point. In panel a) we report the result without SOC while in panel b) the result under the action of SOC. The arrows indicate the points of the Fermi surface four-fold degenerate, where the surfaces are forced by the crystal symmetry to be connected.}  
	\label{FSOC} 
\end{figure}

This produces the two 2D surfaces that are doubly degenerate due to the inversion-time reversal symmetry and are constrained by the crystal symmetry to be connected along the lines where the degeneracy is four-fold.
We emphasise that in this class of compounds a single nonsymmorphic symmetry is not enough to create a Fermi surface with reduced dimensionality, but we need two nonsymmorphic symmetries as above claimed.
Since the two nonsymmorphic symmetries are present also in the non-magnetic phase of CrAs and MnP, it is likely to find two 2D Fermi surface sheets propagating along the $k_{z}$ direction and centred around the SR line. Indeed, from Fig.~\ref{FSCrAs}a)
and ~\ref{FSCrAs}b), we see that also the CrAs shows a three-dimensional surface around the $\Gamma$ point and two 2D FS. Even in this case, there are two holelike 2D surfaces in the $ab$ plane centred around the SR line, but the FS are more cylindrical than those found in the WP.
However, in the magnetic phase, the 2D surfaces around the SR line disappear but other 2D FS appear along other high-symmetry lines of the BZ where one a single nonsymmorphic symmetry takes place.~\cite{Autieri17}\\
\noindent Finally, we investigate the effect of the SOC on the Fermi surfaces of the WP as representative case of this class.
Though the SOC is too small to change the dimensionality of the Fermi surface, it produces selective splittings in the points where the FS is degenerate.
We report the two 2D Fermi surfaces of the WP without SOC in Fig.~\ref{FSOC}a) and including SOC in Fig.~\ref{FSOC}b).
The time-inversion degeneracy is always present.
Without SOC, the two FS are constrained by the nonsymmorphic symmetry to be connected in the four points intersecting the SX and SY lines resulting in a four-fold degeneracy. When the SOC acts, we observe a selective removal of the degeneracy.
In particular, the FS are still forced to be connected on the SY line while the SOC splits them along the SX line.

\section{Discussion}

Concerning the superconducting phases in MX compounds, various considerations are in order. 
\\
The first observation refers to the presence of nonsymmorphic symmetries associated with the glide planes and screw axes within the 
Pnma crystal structure of the targeted compounds. These symmetries can lead to different predictions concerning the nature of the superconducting state and the emergence of non-trivial topological superconductivity.
Indeed, focusing on the case of a glide plane symmetry represented by $g=\{m|${\boldmath$\tau$}$\}$ with $m$ being a mirror operator and {\boldmath$\tau$}  a non-primitive translation operator along a direction in the mirror plane, we observe that, independently of the electronic structure at the Fermi level, there are two basic guiding principles for establishing whether one would end up into a topological nonsymmorphic crystalline superconductor. First, one needs to evaluate whether the superconducting order parameter preserves (even case) or spontaneously breaks (odd case) the glide plane symmetry. Second, one has to assess the relationship of the glide parities between any given electronic state close to the Fermi level and its particle-hole partner that enters into the pairing \cite{Wang16}. Indeed, the particle-hole symmetry can either preserves the glide parity or mixes the parities and such behavior is always opposite for the momentum high symmetry lines in the glide symmetric plane at  ${\textbf{k}}\cdot{\boldsymbol\tau}=0$ and ${\textbf{k}}\cdot{\boldsymbol\tau}=\pi/2$.
This symmetry aspect allows to have a topological invariant only along those one-dimensional cuts and not a global one in the two-dimensional glide invariant plane. As a consequence, depending on the glide parity of the superconducting order parameter, there will be Majorana modes or gap closing topologically protected that occur in correspondence with the projected lines in the momentum space at ${\textbf{k}}\cdot{\boldsymbol\tau}=0$ or ${\textbf{k}}\cdot{\boldsymbol\tau}=\pi/2 $. One can argue that indipendently of the glide symmetry of the superconducting order parameter, for the high symmetry lines in the glide symmetric plane there will always be emerging topological features that can manifest at the edge of the superconductor \cite{Wang16}.
\\
A more specific prediction for the possibility of achieving a topological nonsymmorphic crystalline superconductor can be deduced by considering the topology of the Fermi surface obtained within our {\it ab-initio}  calculations. In this case, benefiting of the construction of the topological invariants performed in Ref. \cite{Daido19}, it is possible to link the $\mathbb{Z}_4$ (glide odd pairing) and $\mathbb{Z}_2$ (glide even pairing) topological invariants for time-reversal symmetric odd-parity superconductors to the number of Fermi surfaces intersecting two distinct connecting paths in the BZ. Our results, in this context, are particularly useful because the knowledge of the Fermi surface allows to give specific expectations on the possibility to achieve non-trivial topological superconducting phases with glide protected surface states. Taking into account the explicit expressions of the $\mathbb{Z}_4$ and $\mathbb{Z}_2$ topological invariants, one needs to evaluate the number of Fermi sheets ($\#FS$) that are encountered along the $R\rightarrow U$ and $S\rightarrow U$ cuts, including the twofold degeneracy due to time and inversion symmetries. In particular, we recall that $\theta^{(a)}_4(\pi)=\#FS_{R \rightarrow U}/2 (\text{mod}\,4)$ and $\theta^{(n)}_4(\pi)=\#FS_{S \rightarrow U}/2 (\text{mod}\,4)$ \cite{Daido19}. With these elements we can summarize the resulting odd parity superconducting phases for the MnP and WP as a function of the filling $n$ for the two glide planes indicated as $a$-glide and $n$-glide \cite{Daido19}.\\

\begin{table}
  \begin{center}
    \caption{We consider the number of Fermi sheets intersecting specific cuts in the BZ and topological invariants for the MnP at different filling $n$ and WP at the nominal electron concentration. 
In the Table we report the number of intersecting Fermi surface along the high symmetry lines ($\#FS_{R{\rightarrow}U}$ and $\#FS_{S{\rightarrow}U}$) 
and the amplitude of the $\mathbb{Z}_4$ topologial invariant \cite{Daido19} corresponding to the $a$-glide ($\theta^{(a)}_4(\pi)$) and 
the $n$-glide symmetry ($\theta^{(n)}_4(\pi)$), respectively.   
Informations about the glide even superconducting phases can be directly deduced by observing that the $\mathbb{Z}_2$ topological invariants are half of the $\mathbb{Z}_4$ ones\cite{Daido19}.}
    \label{tab:table2} 
\begin{tabular}{|c|c|c|c|c|c|}
\hline
& \scriptsize $\#FS_{R{\rightarrow}U}$ &\scriptsize $\theta^{(a)}_4(\pi)$ &\scriptsize $\#FS_{S{\rightarrow}U}$ & \scriptsize $\theta^{(n)}_4(\pi)$ &\scriptsize  2D FS \\
\hline
\scriptsize  MnP with n$>n_1$          & 0     & 0  & 0  & 0  & NO \\
\hline 
\scriptsize  MnP with n$_2<n<n_1$      & 0     & 0  & 8  & 0  & NO \\
\hline
\scriptsize  MnP with n$_3<n<n_2$      & 8     & 0  & 8  & 0  & NO \\
\hline
\scriptsize  MnP with n$_4<n<n_3$      & 8     & 0  & 8  & 0  & YES \\
\hline
\scriptsize  MnP with n$_5<n<n_4$      & 4     & 2  & 4  & 2  & YES \\ 
\hline
\scriptsize  WP  with n=9              & 4     & 2  & 4  & 2  & YES \\ 
     \hline
    \end{tabular}
  \end{center}
\end{table}
\normalsize
The results indicate that for the nominal filling all the candidates odd parity pairing states for the WP compound are nontrivial topological crystalline nonsymmorphic superconductors. As a consequence of this type of topological character and of the bulk-boundary correspondence, one expects to observe surface states protected by the glide appears on the surfaces $(2k,2p+1,0)$ or $(0,2k+1,2p+1)$ ($k$ and $p$ being integers) and with in-gap states having a M\"obius type electronic structure \cite{Daido19}. On the other hand, the MnP compound can be topological only upon doping, because for the nominal filling $n=10$ the eight Fermi surfaces emerging around the S point lead to a zero topological invariant (see the second line of the Table \ref{tab:table2}). Here, we find a direct connection of the multifold degeneracy along the $SR$ line and the potential topological nature of the superconducting phase. Only a variation of the filling can induce a topological transition and according to the Table \ref{tab:table2} it has to be a hole-like doping. The topological region is explicitly highlighted in cyan in the Fig. \ref{BSMnP}. We notice that other energy windows that are much farther from the Fermi level can be also topologically interesting with respect to the nonsymmorphic symmetries.
\\
A final observation concerning the possibility of achieving an unconventional superconducting phase in the MX compounds arises directly from the multifold degeneracy of the electronic dispersion along the SR line. Indeed, if there are Fermi points along the SR line, due to the high degeneracy, one can expect that a high-spin pairing can be realized, in analogy to the unconventional superconductivity proposed in Refs. \cite{brydon16,yang16}. Such observation can be deduced because at any given momentum $k$ close to $k_F$ there are 4-degenerate states for each spin polarization that, thus, realize a quartet for a pseudospin $T=3/2$.  We remind that the configurations associated with the quartet include sublattice and atomic orbital degrees of freedom. Then, combining the angular momentum of the quartet with the spin $s=1/2$, the single particle electronic states at a given $k$ close to $k_F$ can be cast in a configuration with effective total angular momentum $J=L+s$ with amplitude $J=\{1,2\}$. Starting from this consideration, we conclude that the pairing between the particle-hole partners at $\{k,-k \}$ can be obtained by composing single particle configurations with a given $J$ thus building up a total pair angular momentum $J_{pair}$ that can range from 0 to 4. This result implies that the superconducting pairing can be in an effective high-spin state thus going beyond the conventional singlet-triplet classification. Although, the high-spin pairing can be achieved only along the $SR$ high-symmetry lines, we observe that such occurrence would also influence the character of the pairing away from it. Assuming that the superconducting order parameter is smoothly changing across the BZ, we expect that the pairing state evolves in a way that is symmetrically compatible with the presence of a high-spin configuration along the high-symmetry $SR$ line. 
\\
Let us now consider the possible implications of the evaluated electronic structure with respect to the magnetic properties of the MX compounds. 
One direct consequence of the multifold degeneracy is that there are peaks in the density of states, close to the crossing points, which naturally lead, in the presence of interaction, to a tendency to get a Stoner-like instability. On the other hand, according to our results, the fact that a two-dimensional Fermi surface can be observed and that there are almost flat Fermi lines crossing the BZ, which arise from the constraint of the nonsymmorphic symmetries, we expect pronounced nesting effects and thus the possibility of density wave instabilities. 
One relevant observation pertains to the removal of degeneracy in the presence of a magnetic instability. Although there are no induced gap opening, we expect that, in a similar fashion as for the Peierls instability\cite{Park19}, the removal of the degeneracy associated with a magnetic broken symmetry phase leads to an energy gain that stabilizes the corresponding phase. In this context, we observe that a collinear magnetic state can affect the band degeneracy in a twofold way: i) by splitting the states protected by the crystalline nonsymmorphic symmetry, and ii) by inducing new nonsymmorphic invariance and related multiple band crossings\cite{Schoop18a,Schoop18b,Brzezicki17}. More specifically, for the MX materials a collinear antiferromagnetic order has been shown to partially remove the band degeneracy. Depending on the antiferromagnetic pattern, the system can selectively present multifold degenerate bands along inequivalent high symmetry lines\cite{Autieri17}. It is only with a noncollinear magnetic order that one achieves a complete removal of the band degeneracies. We argue that, following the comment on the energy gain by band splitting, it is such mechanism which can account for the observed magnetic order in the CrAs and MnP compounds.
\\
Moreover, it is worth pointing out that the eight-fold degeneracy along the $SR$ line permits to achieve a band insulating phase only for fillings which are multiple of eight contrary to the conventional twofold filling constraint due to the time-inversion degeneracy.
\\
Finally, it is worth mentioning that evidences of spin-triplet superconductivity have been experimentally found for both CrAs and UCoGe\cite{Guo18,Manago19} along the suggested theoretical predictions.
Here, due to the symmetry analogy, our results on the structure of the Fermi surface indicate that unconventional superconductivity can be also observed in WP and MnP with distinct signatures of topological superconductivity manifested by non-trivial zero-energy symmetry protected edge states.
According to our findings, in this framework, the WP could be a topological nonsymmorphic crystalline superconductor while MnP would be a trivial superconductor manifesting a topological transition only upon suitable doping.

\section{Conclusions}
By combining density functional theory with an effective low-energy model Hamiltonian, we have studied the electronic and structural properties of the WP compound in the presence of spin-orbit coupling. The emerging electronic properties are representative of the MnP-type family and also share common features with other members as CrAs and MnP compounds.

Our results indicate that WP exhibits a wider bandwidth and more distorted structure if compared to the CrAs and MnP materials. In all these pnictides the {\it d}-states dominate at the Fermi level while the {\it p}-states are located both above and below the Fermi level. The highly distorted structure, favored by a weak crystal field potential, produces a strong orbital mixing of the {\it d}-bands making quite difficult the identification of a reduced subset of {\it d}-orbitals for an effective model for the low energy description.

The band structure, other than the time-reversal and inversion symmetries, exhibits nonsymmorphic symmetries that bring to four- or eight-fold degeneracy of the bands along some high-symmetry lines of the BZ. Remarkably, nonsymmorphic symmetries have strong implications on the multiplicity and dimensionality of the Fermi surface\cite{Usui11}. Moreover, the evolution of the Fermi surface topology is linked to the presence or not of Fermi points along some high symmetry directions. If the Fermi level crosses the bands along the SR line, there are eight Fermi surfaces with the presence of Fermi pockets and stripes. Depending on the electron filling, the size of Fermi pockets can increase until creating open Fermi surfaces. 
The electronic changeover exhibits a simultaneous modification of the Fermi surface dimensionality and topology of the Fermi surface. 
This picture brings towards four concentric 2D-Fermi surfaces in this class of materials. When the role of the SOC is considered, we show that the interplay between the SOC interaction and the inter-orbital degrees of freedom allows a selective removal of the band degeneracy. We point out that 2D Fermi surfaces were observed also in quasi 2D-superconductors like cuprates,~\cite{Vanharlingen95,Giusti19} ruthenates~\cite{Mackenzie03,Forte10,Autieri12,Malvestuto13,Autieri14,Granata16} and iron-pnictides.~\cite{Si16,Mazin08} However, while in the iron-pnictides the FS are electron-like, in this class of compounds they are hole-like. Furthermore, the transition-metal pnictides MX are intrinsically 3D while cuprates, iron-pnictides and ruthenates may be considered such as quasi 2D systems.~\cite{Porter18} Moreover, the presence of these 2D sheets could affect the transport and superconducting properties of this class of materials. We would like also to stress that the study of nodal excitations could also be relevant when the phonons effects are concerned, allowing to explain the unusual resistivity  behavior~\cite{Nigro18}and possibly shedding light on the mechanism beyond the superconducting pairing.~\cite{Guo18,Manago19}\\

\section*{Acknowledgments}

We would like to thank A. Daido, A. Avella, T. Hyart and K. Sen for useful discussions. The work is supported by the Foundation for Polish Science through the IRA Programme co-financed by EU within SG OP. This research was carried out with the support of the Interdisciplinary Centre for Mathematical and Computational Modelling (ICM) University of Warsaw under Grant No. G73-23 and G75-10.

\section{Appendix}

\subsection{Crystal-field energy levels}

Here we investigate the effect of the octahedral rotations on the crystal-field energy levels. 
This issue is performed by looking at the density of states of WP in three cases. Firstly, we consider the ideal high-symmetry crystal structure as that quoted in the Table~\ref{tab:table1} where there are no octahedral rotations. Then, we consider the experimental volume and the ideal high-symmetry atomic positions where there is the part of the octahedral rotations due to the volume effect. Finally, we examine the experimental structure with the atomic positions calculated from the atomic relaxation procedure with full octahedral rotations.
To estimate the effect of the distortions, we follow the evolution of the $d_{3x^2-r^2}$-energy levels where $x$ is the direction orthogonal to the face-sharing surfaces. 
Without octahedral distortions, these $d_{3x^2-r^2}$ orbital states are the lowest non-degenerate energy levels in the face-shared octahedral crystal field\cite{Khomskii16}. However we show that the energy levels change in presence of the strong octahedral distortions of the WP. Therefore, the $d_{3x^2-r^2}$ states give an indication on how close we are to the undistorted case.

\begin{figure}[t!]
\centering
\includegraphics[height=\columnwidth, angle=270]{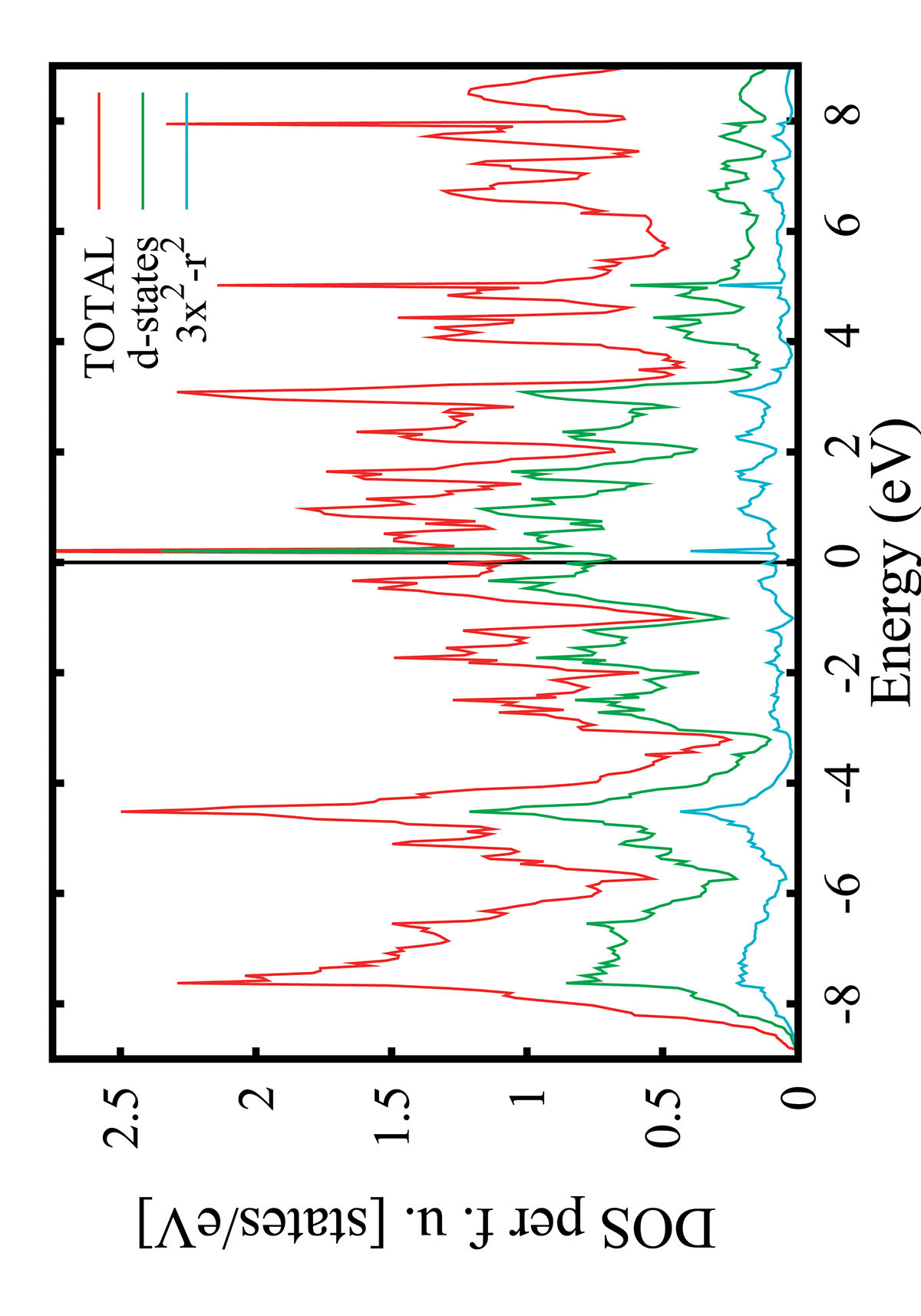}
\caption{DOS for the hypothetical high-symmetry structure. The total DOS per formula unit is plotted as red.
W states are plotted as green and the $3x^{2}-r^{2}$ states are shown in cyan.
The Fermi level is set at the zero energy.}
\label{DOS1}
\end{figure}

\begin{figure}[b!]
\centering
\includegraphics[height=\columnwidth, angle=270]{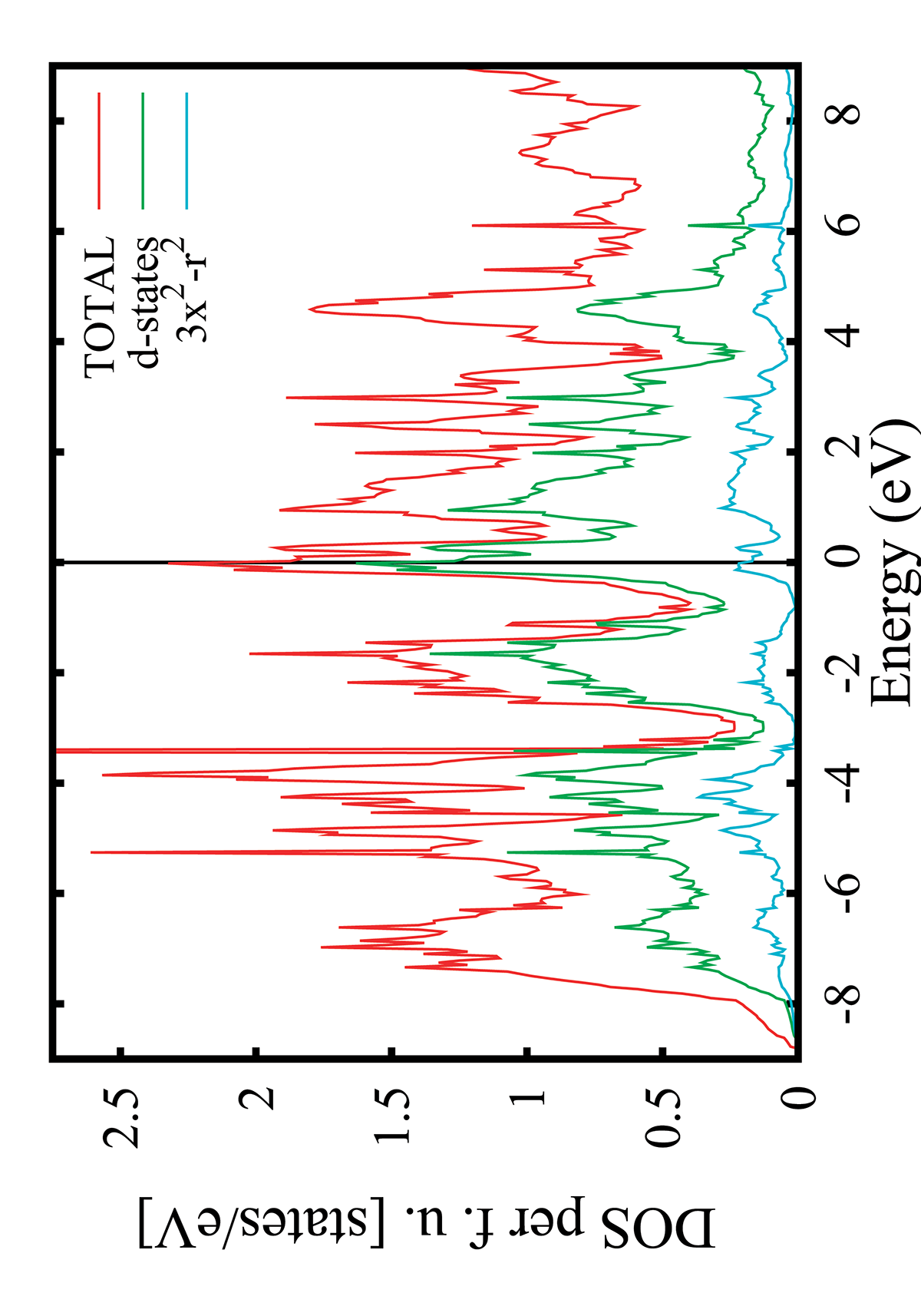}
\caption{DOS of the WP obtained considering the experimental lattice parameters and the high-symmetry atomic positions. The total DOS per formula unit is plotted as red.
W states are plotted as green and the $3x^{2}-r^{2}$ states are shown in cyan.
The Fermi level is set at the zero energy.}
\label{DOS2}
\end{figure}

In the first case, the $3x^{2}-r^{2}$ DOS presents the largest peak at about 5 eV below the Fermi level as we can see from the Fig.~\ref{DOS1}. In the second case, there is a mixing between different energy levels, producing two large peaks approximately located at -4 eV and +2 eV, as it can be observed by inspection of Fig.~\ref{DOS2}. 
In the third case, corresponding to the fully distorted structure, the Fig.~\ref{DOS3} shows that the peak of the $3x^{2}-r^{2}$ DOS is set at about 3 eV above the Fermi level.
Thus, moving from the ideal structure to the fully distorted structure, the $3x^{2}-r^{2}$ energy level peaks move from occupied to unoccupied configurations. One of the reasons why the energy levels change so drastically as a function of the structural properties in this family is the weakness of the crystal field. 
Indeed, in strongly ionic systems, like in transition metal oxides, the transition metal atom is surrounded by atoms with filled p-shells. In these compounds, the transition metal atom is surrounded by atoms with partially filled p-shells, producing a weaker crystal field.
Moreover, the strongly distorted structure, favored by the weak crystal field, produces a mixing of the orbital character of the $d$-bands making impossible to identify a subset of $d$-orbital close to the Fermi level.

\subsection{p-d hybridization in transition metal pnictides}

\begin{figure}
\centering
\includegraphics[width=7cm,angle=0]{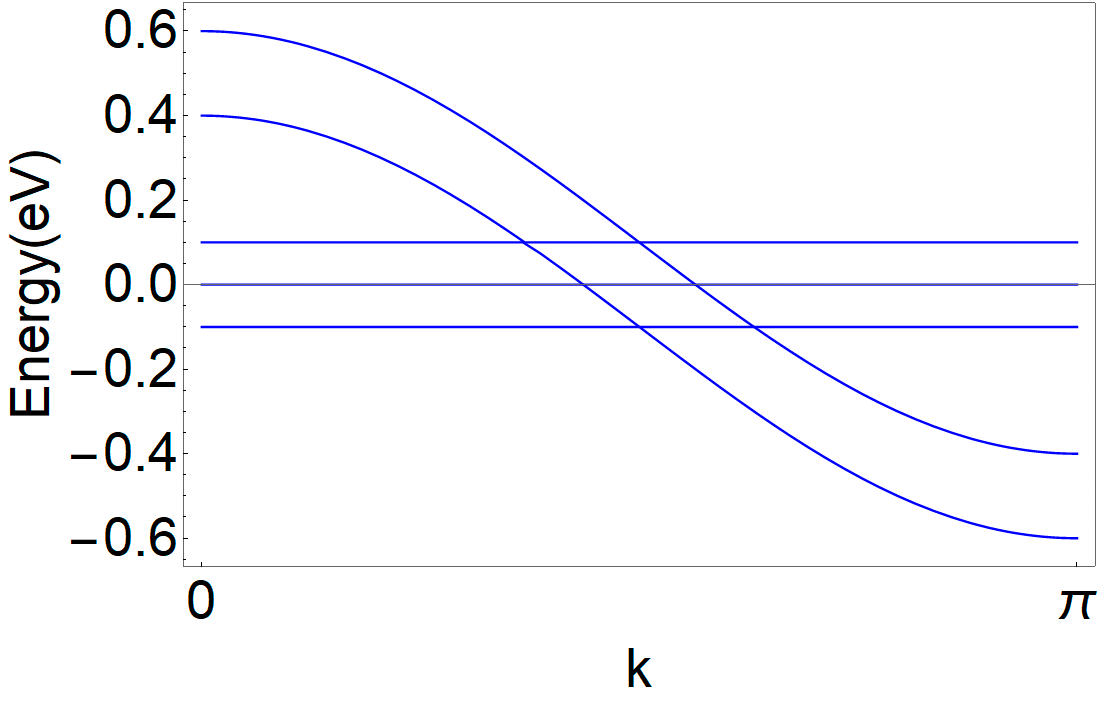}
\caption{Band structure of three flat $d$-bands and two wide $p$-bands when the $d$-$p$ hybridization is turned off. The Fermi level is set at zero energy.
}
\label{dp1}
\end{figure}

\begin{figure}
\centering
\includegraphics[width=7cm,angle=0]{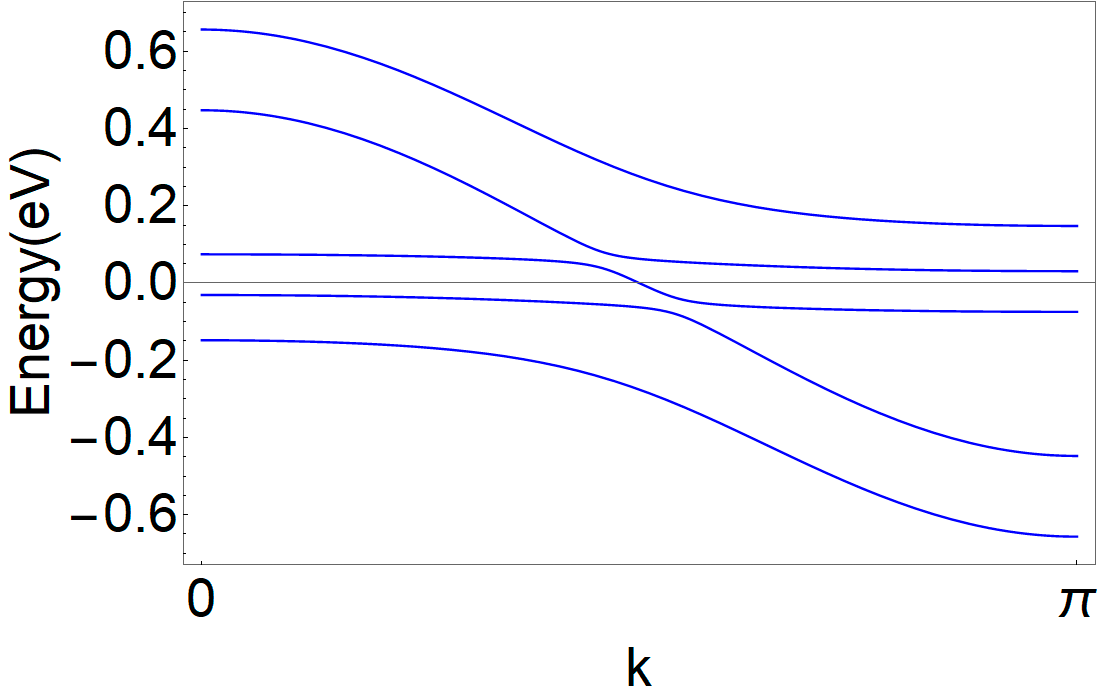}
\caption{Band structure of three flat $d$-bands and two wide $p$-bands when the $d$-$p$ hybridization is switched on. The Fermi level is set at zero energy.
}
\label{dp2}
\end{figure}

Here we examine a $p$-$d$ toy model, in order to demonstrate that it is not possible to reduce to a simple minimal model for the entire subset of $d$-bands close to the Fermi level.
We start by considering three $d$ and two $p$ bands, where the $p$-bands have a large dispersion, while the $d$-bands are flat in the region between -0.1 and 0.1 eV. In Fig.~\ref{dp1} we show the case in which the hybridization between the $d$- and $p$-orbitals is turned off,  while in Fig.~\ref{dp2} we plot the band structure when the hybridization is switched on. From an inspection to these figures, we conclude that one can distinguish the different contribution of the orbitals to the bands in the first case, whereas the same bands are strongly entangled when the $d$-$p$ hybridization is switched on.

This simple example shows that we cannot obtain an accurate minimal model for all three $d$-bands. When the $p$-$d$ hybridization is larger than the difference between the on-site energies of the $p$- and $d$-states, the $d$- and $p$-manifolds cannot be disentangled.\\
However, an effective minimal model can be achieved by limiting to the single $d$-band crossing the Fermi level, as we can see in Fig.~\ref{dp2}. The possibility describe the band structure close to the Fermi level in terms of a minimal set of bands having mainly d character has been recently demonstrated in other transition metal pnictides.~\cite{Cuono19}. We also point out that, when magnetic properties are concerned, the strong hybridizations requires the $p$-channel to be included, like in the Anderson model\cite{Anderson61,Noce06}, thus allowing for a correct interpretation of the magnetism for this class of compounds\cite{Wyso19}.

The impossibility to decouple the $p$-bands from the $d$-bands depends on the covalence and electronegativity of CrAs, WP and MnP not on the group symmetry. Indeed, in other Pnma structure like the perovskite oxide ABO$_3$, it is straightforward to decouple the low energy $d$-bands from all the other bands\cite{Roy15}.

\subsection{Oxidation state}

The oxidation state of CrAs\cite{Autieri17b} and WP is predominantly 0, 
while in MnP the oxidation state is more subtle.
 Indeed, if the oxidation state of Mn had 
been zero, then Mn would have been in 3d$^7$ configuration
with a maximum magnetic moment of 3 $\mu_{B}$. However,
with GGA+U calculations~\cite{Continenza01} and at high volumes,~\cite{Gercsi10} we
can see that the magnetization exceeds 3 $\mu_{B}$ per Mn
atom. Therefore, the Mn oxidation state cannot be zero for the metallic MnP compound. 
We have calculated the oxidation for the non magnetic case obtaining +0.6 for the Mn
and -0.6 for the P. This suggests that also the oxidation state of the Cr can be 
of the order of +0.6 or +1 in CrP.\\

\subsection{Semi-Dirac like energy-momentum at the R point}

At R we have one eigenvalue eightfold degenerate: $E(R)=t_{1}(\pi,\pi,\pi)$.
We can follow the evolution of the eigenvalues from R to U: 
\begin{equation}\label{EigRU}
E_{\pm}(\pi,\pi-\epsilon,\pi)=t_{1}\pm\sqrt{t^2_{3}+ t^2_{4}} .
\end{equation}
The series expansion at first order in $\epsilon$ gives: 
$t_{1}(\pi,\pi-\epsilon,\pi) \simeq t_{1}(\pi,\pi,\pi)$, $t_{3}(\pi,\pi-\epsilon,\pi) \simeq 0$ and 
$t_{4}(\pi,\pi-\epsilon,\pi) \simeq (t^{001}_{AC\alpha\alpha} - t^{00\bar{1}}_{AC\alpha\alpha}) \epsilon$
from which we can conclude that the two eigenvalues 
\begin{equation}\label{EigRU1}
E_{\pm}(\pi,\pi-\epsilon,\pi)=t_{1}(\pi,\pi,\pi)\pm|(t^{001}_{AC\alpha\alpha} - t^{00\bar{1}}_{AC\alpha\alpha})| \epsilon 
\end{equation}
have an opposite and Dirac-like linear behaviour in $\epsilon$ as shown in Fig. \ref{BSMnP}.

\end{document}